\pdfoutput=1
\documentclass[apj]{emulateapj}
\usepackage{amsmath}
\usepackage{amsbsy}
\usepackage{rotating}
\usepackage{color}
\shorttitle{Newly-quenched galaxies and the size evolution of the population}
\shortauthors{Carollo, C.M. et al}

\begin{document}
 
\title{Newly-quenched galaxies as the cause for the apparent evolution in average size of the population}

\author{C.\ M.\ Carollo\altaffilmark{1},  T.\ J.\ Bschorr\altaffilmark{1}, A.\ Renzini\altaffilmark{2}, S.\ J.\ Lilly\altaffilmark{1}, P.\ Capak\altaffilmark{3},
 A.\ Cibinel\altaffilmark{1},   O.\ Ilbert\altaffilmark{4},     M.\ Onodera\altaffilmark{1}, N.\ Scoville\altaffilmark{5}, E.\ Cameron\altaffilmark{1},
 B.\ Mobasher\altaffilmark{6}, D. Sanders\altaffilmark{7}, Y.\ Taniguchi\altaffilmark{8} }
\altaffiltext{1}{Institute for Astronomy, Swiss Federal Institute of Technology (ETH Zurich), CH-8093 Zurich, Switzerland\\ First author email: marcella@phys.ethz.ch}
\altaffiltext{2}{Istituto Nazionale di Astrofisica,Osservatorio Astronomico di Padova, Vicolo
dellÕOsservatorio 5, I-35121 Padova, Italy}
\altaffiltext{3}{Spitzer Science Center, California Institute of Technology, 1200 E. California Blvd, Pasadena, CA, 91125, USA}
\altaffiltext{4}{Laboratoire dÕAstrophysique de Marseille, 38 rue Frederic Joliot Curie, 13388 Marseille, France}
\altaffiltext{5}{California Institute of Technology, MC 105-24, 1200 East California Boulevard, Pasadena, CA 91125}
\altaffiltext{6}{University of California, Department of Physics and Astronomy, Riverside, CA 92508, USA}
\altaffiltext{7}{Institute for Astronomy, University of HawaiÕi, 2680 Woodlawn Dr, Honolulu, HI 96822, USA}
\altaffiltext{8}{Research Center for Space and Cosmic Evolution, Ehime University, Bunkyo-cho 2-5, Matsuyama 790-8577, Japan}

\begin{abstract}
We use the large COSMOS sample of galaxies to study in an internally self-consistent way the change in the number densities of quenched early-type galaxies (Q-ETGs) of a given size over the redshift interval $0.2 < z < 1$ in order to study the claimed size evolution of these galaxies.  In a stellar mass bin at $10^{10.5}<M_{galaxy}<10^{11} M_\odot$, we see no change in the number density of compact Q-ETGs over this redshift range, while in a higher mass bin at $>10^{11} M_\odot$, where we would expect merging to be more significant, we find a small decrease, by $\sim 30\%$. In both mass bins, the increase of the median sizes of Q-ETGs with time is primarily caused by the addition to the size function of larger and more diffuse Q-ETGs. At all masses, compact Q-ETGs become systematically redder towards later epochs, with a $(U-V)$
color difference which is consistent with a passive evolution of their stellar populations, indicating
that they are a stable population that does not appreciably evolve in size. We find furthermore, at all epochs, that the larger Q-ETGs (at least in the lower mass bin) have average rest-frame  colors that are systematically bluer than those of the more compact Q-ETGs, suggesting that the former are indeed younger than the latter.  The idea that new, large, Q-ETGs are responsible for the  observed growth in the median size of the population at a given mass is also supported by analysis of the sizes and number  of the star-forming galaxies that are expected to be the progenitors of the new Q-ETGs over the same period. In the low mass bin, the new Q-ETG appear to have $\sim30\%$ smaller half-light radii than their star-forming progenitors. This is likely due to the fading of their disks after they cease star-formation. Comparison with higher redshifts shows that the median size of  newly-quenched galaxies roughly scales, at constant mass, as $(1+z)^{-1}$. We conclude  that the dominant cause of the size evolution seen in the Q-ETG population  is   that the  average sizes and thus stellar densities  of individual Q-ETGs roughly scale with the average density of the Universe at the time when they were quenched, and that subsequent size changes in individual objects, through merging or other processes, are of secondary importance, especially at masses below $10^{11} M_\odot$.

\end{abstract}

\keywords{Galaxies -- galaxies: formation -- galaxies: evolution.}

\section{Introduction}\label{intro}

The evolution of the median size (i.e., the half-light radius $r_{1/2}$)
of the population of massive ($M_\mathrm{Galaxy}  > 10^{10}M_\odot$) quenched
early-type galaxies (Q-ETGs) at given stellar mass has
been widely highlighted in recent years (e.g., Daddi et al.\
2005; Trujillo et al.\ 2007; McGrath et al.\ 2008; van Dokkum et al.\ 2008;
 Cassata et
al.\ 2011; Szomoru et al.\ 2011; Barro et al.\ 2013; Dullo 
\& Graham 2013; Newman et al.\ 2012; Poggianti et al.\ 2013; Shankar et al.\ 2013,  just to cite a few). The size of the effect is quite large,
with a decrease in median $r_{1/2}$ with increasing redshift
$\propto (1+z)^{-1}$; in coarse terms, this implies that, at a given
stellar mass, the median half-light radius of Q-ETGs is
about a factor of  $\sim 2-3$ smaller at $z \sim  2$ than locally,
corresponding to an increase of over an order of magnitude
in the median mean stellar density within the half-light radius of galaxies.

The wealth of studies quoted above have used
a variety imaging data  taken from space and from the ground,  
at different wavelengths, and have focused on galaxy populations
at different redshifts. Quite naturally, there has
been some debate as to whether obvious observational
biases might have affected the results, such as the possible loss in the noise of
outer, low surface brightness parts of the galaxies, or
the possible effects of color-gradients (e.g., Daddi et al.
2005; Mancini et al. 2010). Younger stellar populations
in the cores of galaxies could result in smaller sizes in
the rest-frame ultraviolet,
where the sizes are often measured, than at the longer
wavelengths, which better sample the stellar mass distribution.

Many of the studies cited above have attempted
to deal with these uncertainties (e.g., Mancini
et al.\ 2010; Szomoru et al.\ 2011), and there is now a
reasonable consensus that there is a real effect to be explained.
For example, there appear to be no strong color
gradients in high-z Q-ETG in those cases in which both
rest-frame UV and optical imaging is available (e.g.,  Toft et al.\ 2007; Guo et al.\ 2011). Thus, there is now general consensus that indeed 
the median size of Q-ETGs is substantially smaller at high redshifts, though apparently 
normally sized Q-ETGs coexist with compact ones, especially among the most massive galaxies (e.g. Saracco, Longhetti \& Gargiulo 2010, 
Mancini et al.\ 2010; see also Onodera et al.\ 2010 for a similar conclusion concerning the velocity dispersion of Q-ETGs at $z=2$).

With some exceptions (e.g., Valentinuzzi et
al.\ 2010; Cassata et al.\ 2011; Newman et al.\ 2012; Poggianti et al.\ 2013) this trend has been often entirely ascribed to
the physical growth of individual galaxies. Rather than a puff-up mechanism, decreasing
the central stellar density of Q-ETGs, the favored picture
has been one in which  Q-ETGs maintain a nearly constant mass
within their innermost few kpcs, and gradually
grow inside-out, building up extended stellar envelopes/halos around such compact,
dense cores (e.g., Cimatti et al.\ 2008; Hopkins et al.\ 2009, 2010;
Taylor et al.\ 2010; Feldmann et al.\ 2010; Szomoru et
al.\ 2011). 
Accretion of small satellites in 
minor gas-poor mergers has been widely entertained as the leading mechanism to grow these stellar envelops (e.g., Naab et al.\ 2009; Hopkins et
al.\ 2009; Feldmann et al.\ 2010; Nipoti et al.\ 2009; Oser et al.\ 2012) and thus increase the radius of  
individual high-$z$ compact Q-ETGs, until they reach their  final $z=0$ dimension.

Yet, this may
be only part of the story, and possibly  a minor one.
First, it is now solidly established
that the population of Q-ETGs has undergone a
strong increase in comoving number density  between $z \sim 2$  and the
present epoch (e.g., Williams et al.\ 2009; Ilbert et al.\
2010, 2013; Dominguez Sanchez et al.\ 2011). The mass functions of different galaxy populations
in Ilbert et al.\ (2010, 2013), based on high quality photometric redshifts
in the   COSMOS field (Scoville et al.\ 2007), indicate
that the number density  of quiescent galaxies has
increased by a factor of  $\sim 2$ since $z \sim 1$, and  by a factor of at least 10 since
$z \sim  2$. These observed number density growth factors for
Q-ETGs match those expected by applying a simple continuity
equation to the time-evolution of the actively star
forming galaxy population (Peng et al.\ 2010,2012). 

In
addition, the analysis of SDSS (York et al.\ 2007) mass functions shows that 
typical passive galaxies with $M< 10^{11}M_\odot$  can only have increased their masses by around 20\% (and
definitely less than 40\%) after quenching (Peng
et al.\  2010). Analysis of the mass functions of SDSS
central and satellite galaxies in Peng et al. (2012) refines
these estimates to an average post-quenching mass
increase of  $\sim 25\%$ for typical central galaxies, and a negligible
increase for satellite galaxies. These constraints
strongly limit the amount of merging that may be  available 
to increase the sizes of galaxies.  This, together with the evolution of the number density, 
can explain why the minor dry merging scenario falls somewhat short from quantitatively accounting 
the observed size growth since $z\sim 2$ (e.g., Oser et al. 2012).

It seems therefore quite likely that the advocated after-quenching growth of individual Q-ETGs contributes only modestly to the observed secular increase of the median size of the Q-ETG population;  given the large increase of the number density of these systems since $z\sim2$, it is plausible that another effect may dominate, namely quenching of star-formation in actively star-forming galaxies that keeps producing, at later epochs, new Q-ETGs with larger size 
than those of galaxies quenched at earlier epochs,  as partly advocated on heuristic evidence by,  e.g., Valentinuzzi et al.\ (2010), Cassata et al.\ (2011), Newman et al.\ (2012) and Poggianti et al.\ (2013). The addition,
at progressively lower redshifts, of progressively larger
Q-ETGs will progressively lower the relative fraction of
the more compact galaxies relative to the total Q-ETG
population, and thus produce an upward evolution of the
size-mass relation. There are good reasons, in an expanding Universe that grows structure hierarchically,  to entertain the notion that later-coming Q-ETGs will be larger
and thus have lower stellar densities than galaxies of similar
stellar mass that are quenched at earlier epochs (e.g., Covington et al. 2011). The apparent  disappearance  of
compact Q-ETGs at later epochs may thus be a false reading
of a reality in which earlier populations of denser Q-ETGs
remains relatively stable in terms of numbers through cosmic time, but become
less and less important, in relative number, at later and
later epochs.

An important question to answer is thus  how the number density of compact Q-ETGs
evolves from high redshifts all the way down to the local Universe.
Searches for local analogs to the compact, massive Q-ETGs observed at
$z \sim 2$  have been undertaken to answer this question, and
have given conflicting results. Specifically, for compact
Q-ETGs in massive galaxy clusters, Valentinuzzi et al.
(2010) have reported evidence for little or no evolution
between $z \sim 0.7$ and $z \sim 0.04$ in this population.
Comparing to the SDSS DR7 \citep{abazajian_et_al_2009}, other studies
have also argued for not much evolution in the number
density of compact Q-ETGs between $z \sim 1.5$ and
the present (e.g., Saracco, Longhetti \& Gargiulo 2010).
Other SDSS studies have however reported a drop of at
least a factor $\sim 20$ between $z \sim 1.6$ and $z = 0.1$ (Cassata
et al. 2011) or even more dramatic than this (e.g., Taylor
et al. 2010). Also, Szomoru et al.\ (2011) find that the
minimum growth in size required to reconcile the size
distribution of quenched  galaxies at $z \sim 2$ with that of
their counterparts at $z = 0$ is a factor $\sim 2$ smaller than
the total median size growth observed in the same redshift interval.

Some of the difference between the apparently conflicting
results may in fact stem from  compactness   having
been quantitatively defined in different ways by different
authors, e.g., either in physical units, or relative to
the average size-stellar mass relation for local Q-ETGs.
Quite often, the {\it fraction} of compact Q-ETGs is considered,
as opposed to their number density.
Other aspects of the analyses need however to come under
scrutiny to reconcile such widely diverse results. For
example, Taylor et al.\ (2010), Cassata et al.\ (2011) and
Szomoru et al.\ (2011) use published SDSS sizes as the
comparison standard at $z = 0$. This is risky, as the input photometric catalogs
may have missed compact galaxies through an imperfect
statistical star-galaxy separation (Scranton et al.
2002). Furthermore, as shown by Cibinel et al.\ (2012) on
the galaxy sample of the Zurich Environmental Survey
(ZENS; Carollo et al.\ 2013a), galaxy sizes smaller than
the seeing Point Spread Function (PSF) are not reliably
recovered from ground-based imaging data. The generally
poor PSF (FWHM well above $\sim 1"$) of the SDSS
images casts therefore doubts as to whether  the published SDSS
galaxy catalogues are adequate for this purpose.
Particularly suggestive is the  number density evolution of compact galaxies presented by Cassata et al.\ (2011), which, based on the self-consistent analysis of GOODS images (Giavalisco et al.\ 2004)  is rather flat from $z\sim 2.5$ down to $z\sim 0.5$, and dramatically drops since $z\sim0.5$, due to the comparison of the GOODS-based datapoint at $z=0.5$ with the SDSS point at $z=0$. Furthermore, the analysis of the number densities in Cassata et al.\ sums up all galaxies above $10^{10}M_\odot$, and thus misses possible differential evolution with stellar mass. 

The present paper seeks to explore this issue  in a carefully
controlled fashion by examining, in the redshift range $0.2<z< 1$, the evolution at constant  size (i.e., half-light radius) of the number densities of Q-ETGs, and of their plausible
star-forming progenitors. We perform our analysis in two bins of stellar mass in which the $I_{814W}<24$ COSMOS sample is complete up to $z=1$, i.e., $10^{10.5}-10^{11}M_\odot$ and
$> 10^{11}M_\odot$; these two bins straddle across the nearly redshift-invariant
characteristic mass $M^* \sim 10^{11}M_\odot$ in the Schechter  (1976)
fit to the mass function of galaxies (e.g., Peng et al.\ 2011
and references therein), enabling us to search for differential effects above and below this mass scale.  A strength of our analysis is
the self-consistent use of data from a single survey, i.e., COSMOS,
thereby avoiding basing our conclusions on 
comparisons between inhomogeneous samples,
and in particular relying on the SDSS data for the low
redshift reference  sample. 
The COSMOS field is ideal for
this study, being unique in having both exquisite photometric redshift
estimates for a very large number of galaxies, based on
deep multi-band photometry, and high-resolution F814W
(I-band) ACS/HST images (Koekemoer et al.\ 2007) over  a large, $\sim$1.8 deg$^2$ area. While
limiting the redshift range between $z = 1$ and $z = 0.2$
restricts the evolutionary lever-arm relative to comparing
higher redshift samples with SDSS catalogues, our
approach has indeed the great advantage that the sizes
of the galaxies can be measured in a uniform way from
a single homogeneous data set of unparalleled statistical
significance.

In our analysis we use aperture measurements for determining
the sizes of the galaxies from the ACS F814W
images because of their higher stability relative to model
fitting approaches when applied to the full morphological diversity
of faint high redshift galaxies. We fully calibrate
however our size measurements against magnitude, size,
ellipticity and concentration biases, and show that, once
both aperture and model-fit measurements are so calibrated,
they well agree with each other, giving us confidence
on their robustness.

In detail, the layout of the paper is as follows. Section \ref{dataset} summarizes the dataset and the basic measurements, and describes the selection criteria for the final galaxy sample in detail. Section \ref{seccor} presents the approach utilized to correct sizes and other structural parameters for systematic biases that affect raw measurements as a function of galaxy magnitude, size, concentration and ellipticity. Section \ref{allresults} presents the redshift evolution of the  number densities at constant size- and surface mass density (i.e., the size and surface mass density functions) for Q-ETGs, and thus our core result, i.e., the constancy of the compact Q-ETG population and the emergence of a newly-quenched population of large ETGs over the $z =  \rightarrow 0.2$ period. Section \ref{discussion} presents the size- and surface mass density functions for star-forming galaxies, and compares  the number densities of the newly-quenched  galaxies with the  number densities of star-forming galaxies, of similar masses and sizes, that are expected  to quench in the $z = 1 \rightarrow 0.2$ interval,  based on  a continuity-equation argument (Peng et al.\ 2010).  This Sections also compares the rest-frame optical colors of  compact and large-size populations of   Q-ETGs, and shows that compact Q-ETGs 
become redder towards later epochs and, at least at masses below $10^{11} M_\odot$, they  are also systematically redder at any epoch, and thus likely older, than corresponding large-size Q-ETGs. This reinforces the interpretation  that the latter are the new comers in the Q-ETG population, which are responsible for increasing the median size of ETGs towards later epochs, without substantial increase in size of individual galaxies. In Section \ref{lastwords}  we conclude. 

Four Appendices present some details of our analysis. Specifically, Appendices \ref{sfrs} and \ref{testscorrs}  provide respectively extra information on the robustness of the measured star formation rates and on the reliability of the corrections that we apply to the structural/size parameters; Appendix  \ref{galfitsizes} highlights the general need to correct such latter parameters even for estimates based on  surface brightness fitting algorithms which take into account the effects of the observational PSF; and  Appendix \ref{peng} finally summarizes the procedure that we follow to  derive   quenching rates   using the prescriptions  of Peng et al.\ (2010).

A cosmological model with $\Omega_\Lambda = 0.7$, $\Omega_M = 0.3$, and $h=0.7$ is adopted, and magnitudes are quoted in the AB system throughout.

\begin{figure}
\epsscale{1.1}
\plotone{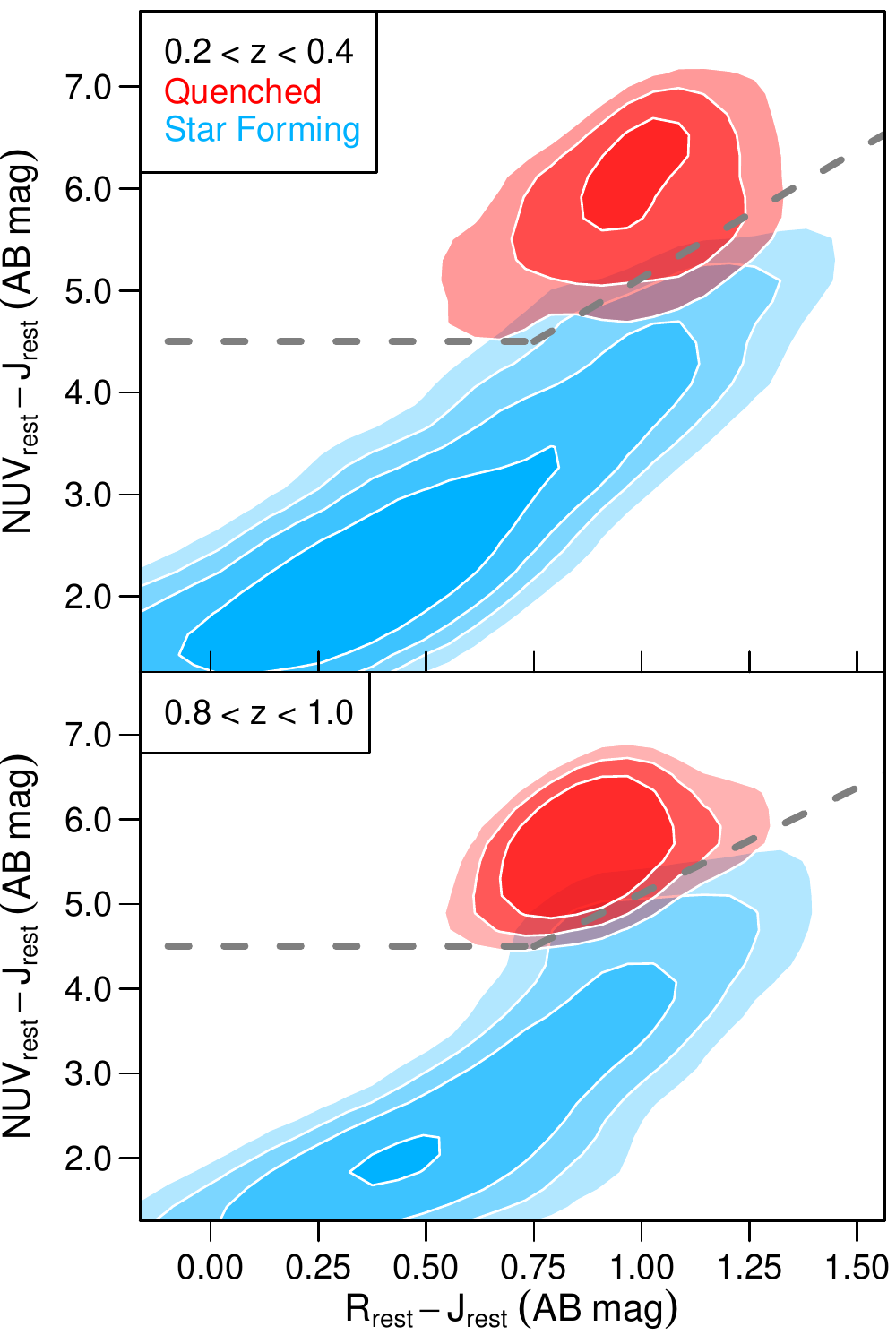}
\caption{The rest-frame NUV$-$$J$ vs.\ $R$$-$$J$ color-color distributions of quenched (in red) and star-forming (in blue) galaxies at $0.2 < z < 0.4$ (top panel) and $0.8 < z < 1.0$ (bottom panel) in the $I_{814W} < 24$ COSMOS sample. Galaxies are classified as quenched or star-forming according to a threshold  in $sSFR=10^{-11}$ yr$^{-1}$. This threshold well agrees with the separation in quenched and star-forming that would be derived using the presented color-color diagram.  To guide the eyes we show as a grey dashed line the saddle line between  star-forming and quenched samples. Quenched galaxies are largely restricted to [NUV$-$$J$ $-$ 2.5$\times$($R$$-$$J$)]$>$ 2.6 mag and NUV$-$$J$ $>$ 4.5 mag.  The shown color contours  are incremented by factors of 2.5, 10, and 25 in number density per bin relative to the baseline, which is a factor of 2 lower in the high redshift bin.\label{f1}}
\end{figure}

\section{The Data and the Basic Measurements}\label{dataset}

\subsection{The COSMOS}\label{cosmossurvey}

We base our study on the COSMOS survey dataset \citep{sco07}, so to capitalize on its high quality compilation of multiwavelength imaging, including Hubble Space Telescope (HST) ACS data \citep{koe07}, over a wide-area field.  For the present analysis we employ the ACS $I$-band source catalogue of \citet{lea07} containing 156,748 sources (102,007 of these tested to be reliable galaxies) down to a flux limit of $I_{814W} = 24$  mag.  The reliability of this catalogue for galaxy photometry and morphological analysis was subsequently improved via extensive visual inspection and cleaning to remove artifacts, cosmic rays, and stars, and to identify deblending errors, leaving a total of 102,007 sources flagged as reliable galaxy detections.

For the purpose of estimating photometric redshifts and stellar masses we match the ACS $I$-band source catalog against the CFHT $i^\ast$-band  \citep{mcc10} and Subaru $i^+$-selected \textit{COSMOS Intermediate and Broad Band Photometry Catalog} (\citealt{ilb09,ilb10}; hereafter, the 'I09 catalog').  This aperture-matched, photometric database, constructed with an updated implementation of the source detection procedure described in \citealt{cap07}, offers a large compilation of broad- and narrow-band flux measurements in 3 arcsec apertures across 31 bands from UV-optical ($u$) through to infrared (8.0 $\mu$m).  As described in Capak et al.\ (2007) the use of PSF-matched, aperture magnitudes allows for a single correction\footnote{The four Spitzer mid-IR IRAC bands \citep{san07} are an exception to this rule as it was unfeasible to degrade the optical data to the much broader Spitzer PSF; rather, the compiled IRAC fluxes were measured in fixed apertures of 1.9 arcsec and are corrected to total by dividing out factors of 0.76, 0.74, 0.62, and 0.58 at 3.6$\mu$m, 4.5$\mu$m, 5.6$\mu$m, and 8.0$\mu$m, respectively; cf.\ \citealt{ilb09}.}  to total flux across all bands for each object; estimates of these corrections are pre-compiled in the I09 catalog and we adopt these for the present analysis.  Two further filter-specific corrections to these total fluxes are then required prior to photometric redshift estimation: \textsc{(i)} a correction against foreground Galactic dust reddening, for which we employ the \citet{sch98} extinction maps with wavelength-dependent adjustment factor, $k_\lambda\times\mathrm{E}(B-V)$, from \citet{car89}; and \textsc{(ii)} a correction against known zero point offsets in the COSMOS photometry, for which we employ the estimates derived from our photometric redshift package {\it ZEBRA}\footnote{The {\it Zurich Extragalactic Bayesian Redshift Analyzer}  (ZEBRA) code is available online at our URL: \texttt{www.astro.ethz.ch/ZEBRA.}} run in catalog-correction mode (Feldmann et al.\ 2006). These corrections are consistent with those published for the COSMOS photometry by \citet{ilb09} and \citet{cap07}; they are based on $\chi^2$-minimization of fitting errors between the best-fit SED template and observed fluxes of galaxies with known redshifts from the zCOSMOS 20k sample \citep{lil07,lil09} (see Feldmann et al.\ 2006).

For book-keeping purposes we note that our procedure for matching the HST/ACS $I$-band source catalog against the I09 multiwavelength photometry catalog (2,017,800 sources) using a 0.6 arcsec tolerance on the centroid offset yields a total of 94,908 (93.0\%) direct galaxy matches (i.e., unique galaxy-galaxy associations).  A further 5267 (5.2\%) galaxies were identified as sharing their match in I09 with another object in the HST/ACS catalog;  based on our visual inspection of a few hundred such systems drawn randomly from the sample, these second matches are typically  neighboring 'junk-sources' or  over-deblended fragments of the original matched galaxy.  We thus treat as successful matches the 3614 of these 5267 duplicates for which the primary match has a smaller centroid offset against its I09 counterpart than the additional (junk) candidate match.  Conversely, due to the relatively broad PSF of the ground-based imaging used in construction of the I09 catalog, only a small number (16, i.e., $\sim0.02$\%) of galaxies in the HST/ACS-based sample were matched against more than one possible I09 counterpart within our 0.6 tolerance; once again, after visual inspection of these few sources, we adopted as the valid matches the sources displaying the smallest offset between the HST and I09 centroids.  A total of 98,538 galaxies were thus deemed successfully matched, leaving only 1816 (1.8\%) galaxies in the parent HST/ACS catalog  unmatched to any object in I09.

One further issue to deal with in the so-obtained galaxy sample is the large degree of flux contamination from interloper objects within the 3$"$ adopted aperture in the I09 catalog. A total of 13,025 (13\%, of our successful matches) were flagged by \citet{cap07} and \citet{ilb09} as suffering severe contamination from either brighter neighbors or from the diffraction spikes of over-exposed stars in at least one of the Subaru $B_J$, $V_J$, $i^{+}$, or $z^{+}$ filters.  Using such flagged objects may introduce errors in our scientific analysis, and hence we exclude these systems from our study.  The completeness of our final galaxy sample is consequently  $\sim$84\% (85,513/102,007), contributing a level of uncertainty to the absolute normalization of the size- and surface mass density $\Sigma_\mathrm{MASS}$-functions at each epoch computed herein comparable to that induced by cosmic variance \citep{tre08} in the COSMOS field \citep{oes10}.  We checked that the completeness of our galaxy sample does not vary markedly with either size (i.e., half light radius) or concentration index, so we do not expect any  size- or morphology-dependent biases in the presented analysis.

\subsection{Photometric Redshifts, Stellar Masses \& Star Formation Rates}\label{stellarmasses}

We estimate photometric redshifts for objects in our matched source catalog using our  ZEBRA code \citep{fel06}. 
Calibration of the benchmark SED templates \citep{col80,kin96} employed in this analysis was achieved by comparison against a sample of $\sim$20,000 galaxies with secure spectroscopic redshifts from the zCOSMOS survey \citep{lil07,lil09}. 
Only 236 of the 85,513 input matched galaxies  were found to be outliers (i.e., a 0.3\% failure rate).  By comparison against the zCOSMOS sample at $I_{814W} < 22.5$ mag, we estimate a photometric redshift uncertainty of $\Delta(z)/(1+z)\sim0.007(1+z)$ at this brightness level.   The uncertainty for  galaxies down to $I_{814W} = 24$ mag was estimated to be $0.012(1+z)$ by artificially dimming the photometry of the zCOSMOS reference sample. The statistical quality of the ZEBRA  photo-z's is very similar to that of the COSMOS photo-$z$ catalog of Ilbert et al.\ (2009). The latter was used to further  validate the robustness of our results towards systematics uncertainties in the photo-$z$'s. 

For each galaxy for which a photometric redshift estimate could be obtained, we further employed a non-public extension of ZEBRA (i.e., "ZEBRA+", see Oesch et al.\ 2010) to estimate the corresponding star formation rates (SFRs) and  stellar masses   based on synthetic SED fitting  to 11 photometric broad--bands, ranging from $3832$\AA\   ($u^\ast$, CFHT) to $4.5\mu$m (Spitzer IRAC channel 2)\footnote{Note that our stellar masses are defined as the integral of the SFR; they are thus about 0.2 dex larger than stellar mass computations which subtract the mass-return from stellar evolution to the interstellar medium.}. The SED library consists of a comprehensive set of  star formation history models,  i.e.,  exponentially-declining SFRs spanning a range of metallicities from 0.05 to 2 $Z_\odot$, decay timescales from $\tau \sim 0.05$ to 9 Gyr, and ages from 0.01 to 12 Gyr (with a Bayesian prior to bound the latter at less than the age of the Universe at any given redshift). 
The construction of this library was achieved via the \citet{bru03} stellar population synthesis code, adopting a Chabrier IMF \citep{cha03}.  The impact of dust extinction is handled during template matching by allowing dust reddening (Calzetti et al.\ 2000) with the E$(B-V)$ value treated as a free parameter of the fit. 
Synthetic template matches were identified, and a stellar mass successfully derived, for all but 1,088 of the 85,277 COSMOS galaxies with photometric redshifts (a 1.3\% failure rate). Owing to the inherent degeneracies in the choice of stellar population template and dust extinction model, which dominate the error budget in the present analysis (given the thorough characterization of the observational SEDs across our multiwavelength database) we estimate an uncertainty of $\sigma_{\log M} \approx 0.20$ dex on our model stellar masses.

For the purposes of separating star-forming  from  quenched  galaxies, we adopt a subdivision at an SED-fit specific star formation rate (i.e., SFR per unit stellar mass; hereafter sSFR) of $10^{-11}$ yr$^{-1}$; this corresponds closely to the inverse age of the Universe at $z \sim 0.3$, the midpoint of the low redshift bin used in our  analysis. In Appendix \ref{sfrs} we explain our choice to use star formation rates based on SED fits rather than (available) IR/UV-derived estimates.  In Fig.\ \ref{f1} we show, on  the rest-frame $NUV-J$ vs.\ $R-J$ diagram, the distributions of  quenched and star-forming classified galaxies in the  $0.2 < z < 0.4$ and $0.8 < z < 1.0$ redshift bins, respectively; star-forming and quenched galaxies are known to effectively separate in different regions  in this diagnostic plane (e.g.\ Williams et al.\ 2009, \citealt{bra09,bun10}).  Inspection of the figure offers confidence that our chosen cut in sSFR well separates star-forming from quenched galaxies in our sample.

\subsection{Morphological Classification}\label{morphologicalclassification}

The morphological classifications  were derived with the {\it Zurich Estimator of Structural Types Plus} ({\it ZEST+}), an upgraded version of the {\it ZEST} approach described in detail in Scarlata et al.\ (2007a).  {\it ZEST+} implements well-tested, robust algorithms for measuring a variety of non-parametric indices, including concentration, asymmetry, clumpiness-, Gini and M$_{20}$ coefficients  for a quantitative structural analysis and morphological classification of faint distant galaxies (see also references in Scarlata et al.\ 2007a).  
The new version of the algorithm, {\it ZEST+},  features several substantial improvements relative to {\it ZEST} in key computations, including a quality-controlled removal of contaminating sources through substitution of optimal sky-valued pixels, a more robust identification of the galaxy centers, important especially for computations of asymmetry and concentration parameters, and  a more robust calculation of the sources'  Petrosian radii, unaffected by noise and contaminations by nearby sources.
{\it ZEST+} also implements a Support Vector Machine (SVM) approach to estimate galaxy morphologies, in addition to the Principal Component Analysis (PCA)  of the previous {\it ZEST} version;  the  morphological classiÞcation of the COSMOS sample that we use in this work uses the SVM approach. 

Both SVM and PCA method require a training sample to guide the morphological classification. The adopted training set is classified in three main morphological types, i.e.,  "early-type" (E/S0) galaxies , "disk" (Sa to Scd) galaxies, and "very late-type" (Sd/Irr/Pec) galaxies. The galaxies in the training set were  carefully selected as archetypal examples of their classes, well separated from non-class members in (at least) concentration, asymmetry, and Gini coefficient, and spanning a representative range of ellipticities, sizes, and apparent magnitudes (corrected for biases, as discussed in Section \ref{sizecorrections}).  The morphological classification was performed on the reduced HST/ACS $I_{814W}$ frames.  Visual inspection of the {\it ZEST+} classified sample, and a quantitative inspection based on simulated images,  confirms a relatively small ($<15\%$)  incidence of catastrophic failures in the classification, down to and a much higher frequency of correctly identified morphologies than in our earlier classification attempt using the original version of {\it ZEST} \citep{cam10}. 

We emphasize that our main findings are not affected by the choice to add a  morphological selection to the samples of quenched and star-forming galaxies.  We indeed checked that all results stand qualitatively unchanged by removing all morphological constraints, with only minor  quantitative differences which do not affect our main conclusions (see also Section \ref{conteq} for a further remark on this point).

\subsection{The Final Galaxy Sample}\label{sampleselection}

The final selection criteria applied to construct the master sample of massive galaxies  used in the present analysis were as follows: \textsc{(i)} {\it ZEBRA} maximum likelihood photometric redshift in the interval $0.2 < z < 1.0$ (see Section \ref{stellarmasses}); \textsc{(ii)} {\it ZEBRA+} maximum likelihood stellar mass greater than $10^{10.5}$$M_\odot$ (see again Section \ref{stellarmasses}); \textsc{(iii)}{\it  ZEST+ morphological type} corresponding to E/S0 and  Sa-Scd galaxies (see Section \ref{morphologicalclassification} above); \textsc{(iv)} no excessive flux contamination in the groundbased CFHT/Subaru imaging from neighboring objects (as indicated by Capak et al.'s 'bad photometry' flag; see Section \ref{cosmossurvey}); and \textsc{(v)} no contamination by artifacts, cosmic rays, neighboring stars, or from deblending errors in the HST/ACS $I_{814W}$ imaging (see again Section \ref{cosmossurvey}). Using these selection criteria,  and thanks to the  excellent combination of  relatively deep ACS images over a $\sim1.8$deg$^2$ field, COSMOS returns a final sample of 11,311 (5355 quenched) galaxies, split in redshifts as follows: 1743 (921) at $0.2 < z < 0.4$, 1751 (833) at $0.4 < z < 0.6$, 3093 (1566) at $0.6 < z < 0.8$, and 4724 (2035) at $0.8 < z < 1.0$.  The variations of numbers of galaxies in the redshift bins highlights that, even COSMOS, is not unaffected by cosmic variance. This is a well-known fact for this field, and needs attention in order to perform studies that involve the evolution of the number densities of sources of a given kind. Following the approach that we also adopted in Oesch et al.\ (2010), we correct for cosmic variance issues as discussed in Section \ref{allresults}.

Finally, we note that, in this paper, we define as `Q-ETGs'  galaxies that are quenched, according to our sSFR-based definition of Section \ref{stellarmasses}, and have an early-type morphology, according to our classification of Section \ref{morphologicalclassification}. 

\section{Galaxy Sizes: Biases and Correction Functions}\label{seccor}

\subsection{Raw Size Measurements}\label{rawsizes}

As indicated above, and as also done in the literature quoted above, we adopt the half-light radius $r_{1/2}$ as a measure for  the  size  of galaxies in our sample. Whilst size measurements based on  aperture fluxes are well-known to prone to systematic biases in the small-size and low surface brightness regimes (e.g.\ \citealt{gra05,cam07}), we nevertheless favor this technique over a profile-fitting approach (cf.\ \citealt{sar07}, Mancini et al.\ 2010, Cassata et al.\ 2011, Whitaker et al.\ 2011) for the present analysis due to its high stability---both the stability of its performance across the full morphological diversity of the high redshift galaxy population, which becomes increasingly irregular/clumpy and problematic for model fitting codes \citep{elm09,oes10}, and the stability (i.e., predictability) of its systematic biases, which can thus be robustly corrected, as we demonstrate in Section \ref{sizecorrections} below.  Conversely, profile fit-based sizes are themselves prone to increasingly unstable behavior (i.e., large random errors) in the low surface brightness regime \citep{hau07} and at faint magnitudes; the results of profile-fits are unreliable for typical galaxies fainter than $I_{814W} \sim 23$ mag in the COSMOS imaging, whereas we want to push our limit one magnitude fainter. 

To measure the (aperture) sizes of the  galaxies in our sample, we used our  custom-built software package package {\it ZEST+} (see Section \ref{morphologicalclassification}). This also measures galaxy half light radii within elliptical apertures, providing an alternative to the circular aperture measurements of SExtractor (Bertin \& Arnouts 1996), which are intrinsically limited in their ability to properly characterize elongated systems such as inclined disks.  Specifically, given an input measurement of each galaxy's apparent total flux (we employ the 2.5$R_\mathrm{Kron}$ value from SExtractor; Kron 1980) and a catalog file highlighting the positions of neighboring objects (again from SExtractor), {\it ZEST+} estimates the local sky background of the galaxy at hand, after replacing the companion galaxies segmentation maps with random sky values, and returns an estimate for the semi-major axis of the corresponding elliptical aperture enclosing half this total flux.

\begin{figure*}
\epsscale{1.1}
\plotone{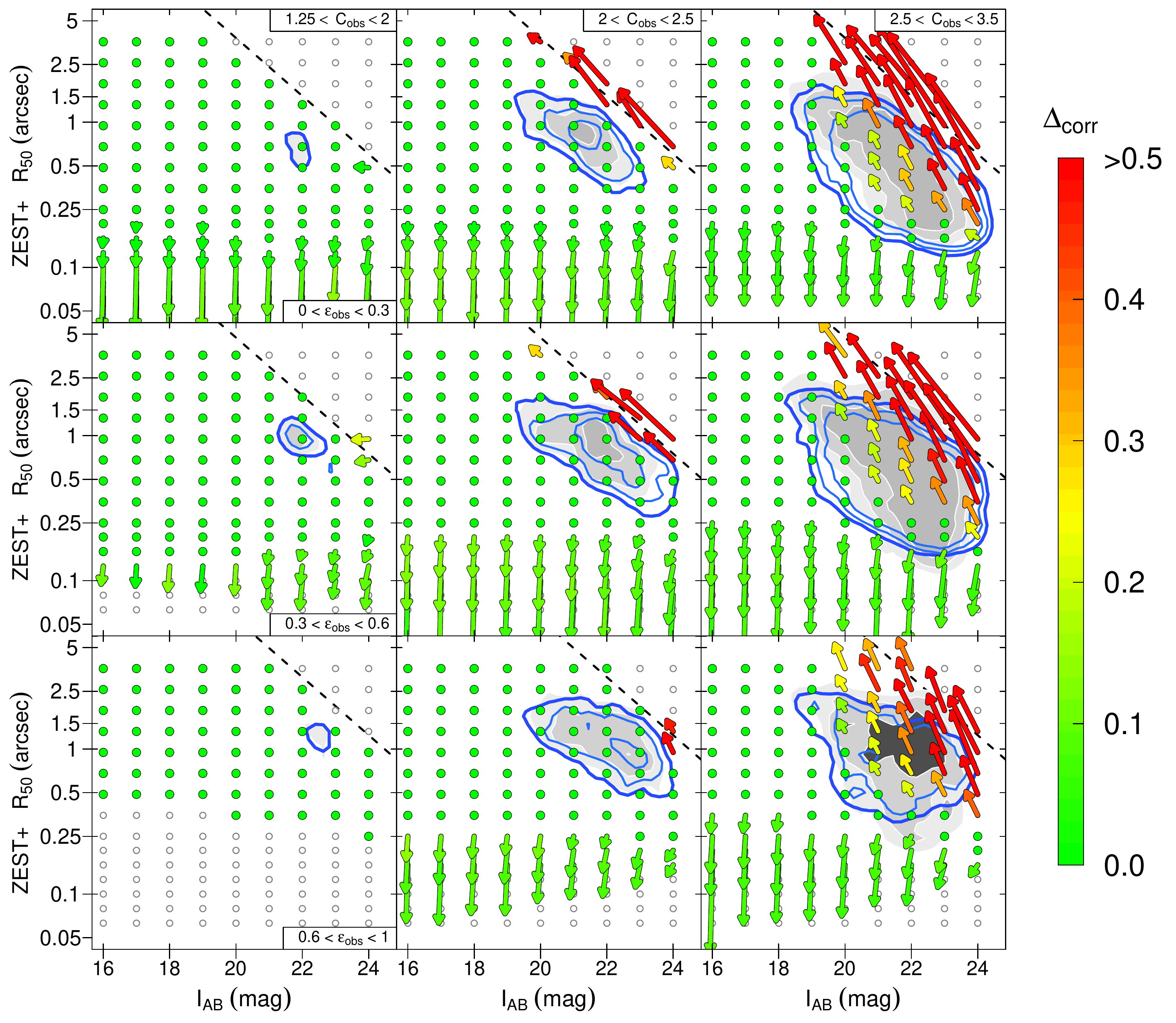}
\caption{Grid of {\it observed} half-light radius and magnitude values, obtained with the aperture approach  (ZEST+), showing the corrections vectors (represented by the arrows) required  at each point of the grid to eliminate systematic biases against these same parameters. The correction vectors  specifically give the extent and direction of the  average corrections to sizes and magnitudes required at each specific point of the (observed) size-magnitude grid.
The nine different panels emphasize the strong influence of profile steepness, and the milder but significant influence of profile ellipticity, in shaping the observational biases. In particular, the  correction vectors are shown in three bins of observed concentration index (left to right: $1.25 < c_{\mathrm{obs}} < 2$, $2 < c_{\mathrm{obs}} < 2.5$, and $2.5 < c_{\mathrm{obs}} < 3.5$), and in three bins of observed ellipticity top to bottom: $0 < e_{\mathrm{obs}}  < 0.3$, $0.3 < e_{\mathrm{obs}}  < 0.6$ and $0.6 < e_{\mathrm{obs}}  < 1.0$).  Green circles indicate points in the observed size-magnitude grid in which the corrections are $<0.3$ magnitudes and $<25\%$ in radius. The colors of the arrows indicate the degree of uncertainty in the amplitude or direction of the arrows, with a scale from green (negligible errors) to red ($>50\%$ error in recovery of the amplitude or direction of the correction). Grey open circles indicate regions of the observed parameter space not populated by enough 'observed' artificial galaxies to apply a reliable correction (indicating that galaxies which may populate these grid points would be scattered out, by observational biasses, into other regions of parameter space; see also Fig.\ \ref{f3} and related discussion in the text). The blue contour lines and the grey-shaded contours in each panel show respectively the distributions of galaxies in our final $0.2 < z < 1.0$ COSMOS sample  for uncorrected and corrected measurements (in steps of a factor of two in number density per contour). The dashed line in the top-right corner of the panels shows the surface brightness limit for the $I_{814W}<24$ COSMOS sample. \label{f2}}
\end{figure*}

\subsection{Correction Functions for Galaxy Sizes: Dependence on Size, Magnitude, Concentration and Ellipticity}\label{sizecorrections}

It is well established that the brightness of the background sky and the blurring effect of the PSF have the potential to introduce size and surface brightness-dependent biases in both the completeness function of galaxy detection and in the recovery of key morphological/structural parameters (cf.\ \citealt{cam07,sar07}); hence, the assessment and, if possible, correction of any such biases is an essential requirement for the robust investigation of the evolution of galaxy sizes.  Despite this general awareness, studies of galaxy sizes typically limit themselves at quoting global  underestimates/overestimates   of sizes by a certain fractional value in different regimes of brightness and radial decline of stellar densities (the latter typically represented by Sersic (1963) indices ranging between $n=1$ and $n>4$, which more or less bracket the  surface brightness radial declines of exponential disks and elliptical galaxies at $z=0$). Yet the systematic uncertainties may depend also on other parameters, most noticeably the ellipticity and size itself of the stellar structure. Here we use the  approach that we introduced in Cibinel et al.\ (2013) for the ZENS galaxy sample (Carollo et al.\ 2013a), and apply the appropriate corrections to measured galaxy sizes as a function of galaxy magnitude, size itself, concentration, and ellipticity. 

In order to  correct for systematic biases in the recovery,  from COSMOS ACS $I_{814W}$ imaging, of galaxy flux and shape parameters, we performed $>2,000,000$ artificial galaxy simulations using Galfit \citep{pen02}.  The surface brightness of each model galaxy was defined via an elliptical-isophote, S\'ersic profile with intrinsic (input) properties drawn randomly from the following parameter space: $14 < I_{814W} < 25$ mag, $0.03 < r_{1/2} < 6$ arcsec, $0 < e < 0.95$, and $0.2 < n < 8$.  Each model galaxy image was constructed at a pixel scale of 0.03 arcsec, matched to our COSMOS/ACS data,  and convolved with a representative ACS $I_{814W}$ PSF \citep{rho07}.  The model images were then added with Poisson noise to empty regions of sky from the real COSMOS HST/ACS images.  

Object extraction was performed on each artificial galaxy using an identical approach to that employed in constructing the COSMOS $I_{814W} < 24$ mag source catalog.  That is, we ran SExtractor first in a ``bright source mode'' (with control parameters: \texttt{detect\_minarea} $=$ 140, \texttt{detect\_thresh} $=$ 2.2, and \texttt{back\_size} $=$ 400), and in the event of null detection a second run was initiated in ``faint source mode'' (with control parameters: \texttt{detect\_minarea} $=$ 18, \texttt{detect\_thresh} $=$ 1.0, and \texttt{back\_size} $=$ 100).  The relevant photometric and structural parameters output by SExtractor (i.e., the Kron magnitude, Kron half light radius, and isophotal ellipticity) were recorded for each artificial galaxy thus extracted.

The simulation was then advanced to the {\it ZEST+} analysis stage: as noted earlier, we  base our analysis on the elliptical aperture galaxy sizes provided by {\it ZEST+}, rather than circular aperture measurements from SExtractor.  Hence, as the final stage of our simulation process, each artificial galaxy detected above the sky noise was  run through {\it ZEST+} to quantify its concentration index, ellipticity, and half light radius in a manner consistent with our treatment of the real COSMOS HST/ACS sources.

The suite of artificial galaxy simulations thus compiled allows to establish the impact of observational biases on the measurement of photometric and structural parameters of galaxies in the COSMOS $I_{814W}$ HST/ACS catalogue by using the output SExtractor and {\it ZEST+}  measurements  to determine {\it correction functions}, for all relevant parameters,  at each point of a dense grid in the four-dimensional space of {\it observed} magnitude, size, concentration and ellipticity. Specifically, starting from regularly-spaced points on a grid in this four-dimensional {\it observed} space, limited by the above boundaries, we selected all models with (output/observed)  magnitude within 1 mag ($\Delta mag =1$),  sizes within  $\Delta r=0.015"$ (or $\Delta r = 0.03''$) for effective radii larger than $0.2''$), ellipticity within $\Delta\epsilon=0.2$ and concentration  within $\Delta C=0.5$  of the targeted point on the observed grid. We then computed the median values of the input model parameters  that had generated these output (observed)  values, and  derived correction vectors connecting the targeted point on the output grid with the original point in the input grid, i.e., the point with coordinates equal to the median of the input values. A correction vector  for each individual galaxy was then obtained by interpolating the correction vectors derived for the grid points.

The  resulting correction functions  are illustrated in Fig.\ \ref{f2}, which shows the grid points in the observed radius versus magnitude diagrams, in three bins of concentration (left to right) and ellipticity (top to bottom).  The length and direction of the arrows give the point-by-point {\it median} correction vectors. The color of the vectors 
is coded according to the  scatter  $\Delta_{corr}$ (in amplitude and/or direction) measured  around the median  of the  correction vector (plotted). The scatter is defined as $\Delta_{corr} = \sqrt{[\sigma(r_{1/2})/<r_{1/2}>]^2+[\sigma(mag)/<mag>]^2}$, with 
$\sigma(r_{1/2})$ and $<r_{1/2}>$ and $\sigma(mag)$ and $<mag>$
 respectively the 1-$\sigma$ scatter values and  medians of the distributions of recovered half-light radii and magnitudes at the given grid point, respectively. The color of the vectors varies from green, for $\Delta_{corr} < 15\%$, up to red, for $\Delta_{corr} > 50\%$.

A few well known trends have been thoroughly commented on in previous studies  (e.g.\ \citealt{sar07,tru07}), e.g.,
 the over-estimation of sizes for objects with half light radii $\leq FWHM$ of the PSF,  and the under-estimation of sizes (and fluxes) for faint, low surface brightness systems with high concentration indices (Mancini et al.\ 2010). Inspection of the figure reveals that the required correction functions  depend on all four parameters (size, magnitude, concentration and ellipticity). Applying these corrections to the raw measurements is important in order not to introduce systematic biases in the measurements.
In this work we make the  step of  correcting the observed parameters for each observed galaxy in our COSMOS/ACS catalogue,  by statistically  recovering the  intrinsic 'true' parameters through interpolation between the  vectors in the four--dimensional  calibration grid discussed above. 

\begin{figure*}[t!]
\epsscale{1.1}
\plotone{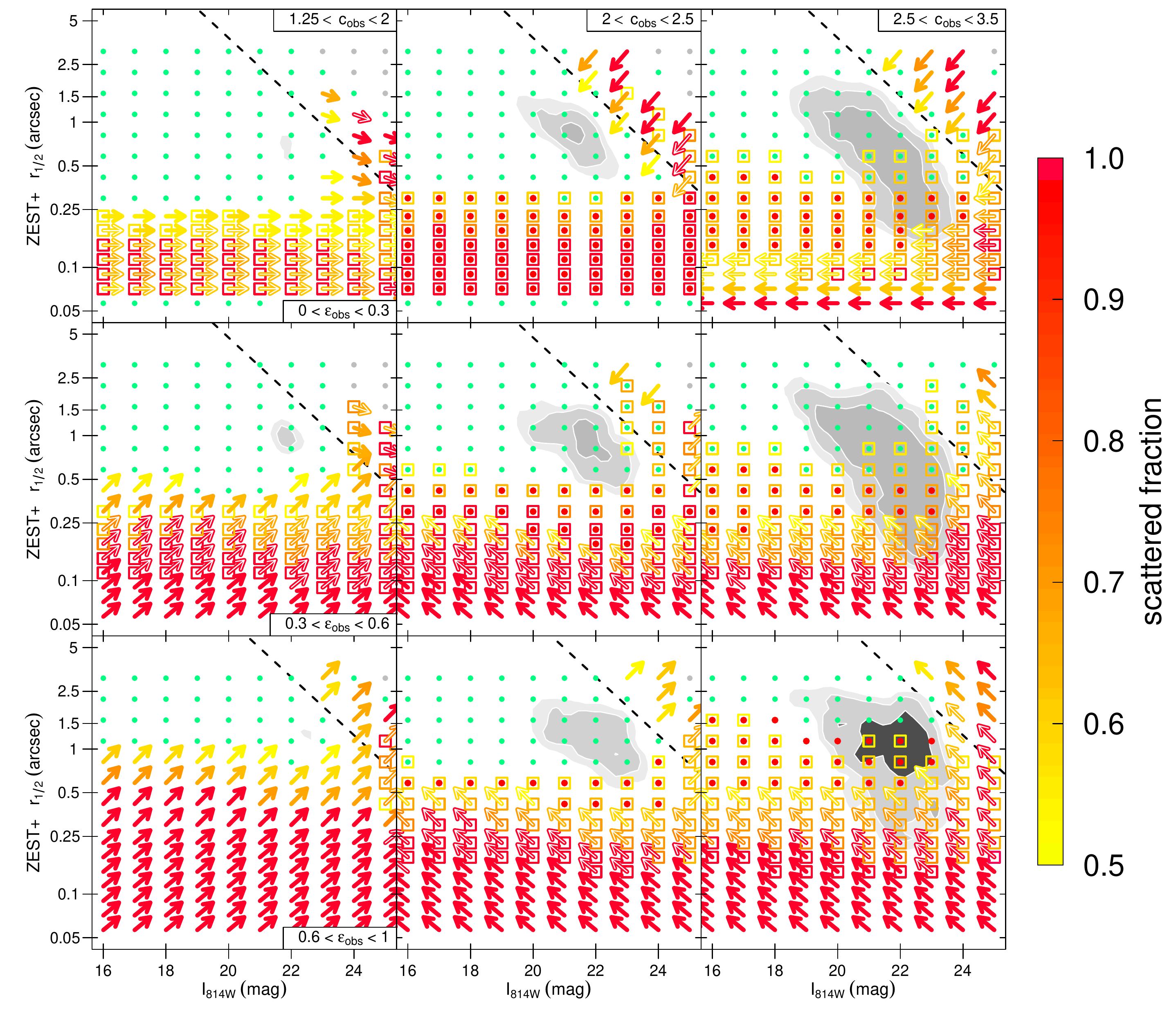}
\caption{Grid of {\it true/input} half-light radius and magnitude values, obtained with the aperture approach  (ZEST+), shown as in Fig.\ \ref{f2} split in three bins of observed concentration index (left to right: $1.25 < c_{\mathrm{obs}} < 2$, $2 < c_{\mathrm{obs}} < 2.5$, and $2.5 < c_{\mathrm{obs}} < 3.5$), and in three bins of observed ellipticity (top to bottom: $0 < e_{\mathrm{obs}}  < 0.3$, $0.3 < e_{\mathrm{obs}}  < 0.6$ and $0.6 < e_{\mathrm{obs}}  < 1.0$).  
Filled green circles  show grid points in which more than $50\%$ of model galaxies born with these given input concentration $C$ and ellipricity $\epsilon$ parameters, are correctly recovered with a negligible scatter  relative to these input parameters (namely, $\Delta C < 0.3$ and $\Delta \epsilon < 0.2$). Filled red circles  show grid points in which $>50\%$ of model galaxies simulated with these parameters is recovered within the original broad bin of concentration and ellipticity (i.e., same panel), but with a large scatter in concentration index ($\Delta C > 0.3$) and in ellipticity ($\Delta \epsilon > 0.2$) relative to the true input parameters. Squares show  grid points  in which $>50\%$ (yellow, up to $>90\%$, red)  of model galaxies are detected at that grid-point but originate from a different bin (panel) of concentration or ellipticity.  Solid arrows show grid-points from which at least $50\%$ (yellow, up to $>90\%$, red) of model galaxies  born  there are scattered out from the given ellipticity/concentration bin (i.e., panel)  and  are 'observed' in a different panel of ellipticity and concentration. Empty arrows show grid-points from which similarly color-codes fractions of galaxies which are born with those given ellipticity/concentration values remain within the same ellipticity/concentration panel bin, but are recovered with errors in concentration $\Delta C> 0.3$ and ellipticity $\Delta \epsilon > 0.2$, respectively. The direction of the arrows visualizes the global 'scattering direction', due to measurement errors,  of galaxy models out of that bin. In particular,  left/right-ward oriented horizontal arrows indicate recovery-errors in concentration only, and specifically scattering towards lower/higher concentrations, respectively. The 45$^o$-inclined left/right- and up/down-ward directed arrows indicate recovery errors in both concentration and ellipticity, and precisely towards lower/higher concentrations and lower/higher ellipticities, respectively. Remaining symbols in the figure are as in Fig.\ \ref{f2}.
\label{f3}}
\end{figure*}

 In Fig.\ \ref{f3} we show, on the same size-magnitude grid points and fixed concentration/ellipticity panels ("bins"), how systematic measurement errors in concentration and ellipticity scatter model galaxies out of their input/true bins  and into incorrect bins of these parameters. The main point of the figure is to give a "glimpse impression" of problematic (yellow to red) regions in the four-dimensional parameter space of magnitude, size, concentration and ellipticity. In detail, filled green circles in the panels indicate grid points in which model galaxies simulated with these parameters are more than $50\%$ of the times correctly recovered within the original bin of concentration and ellipticity, and with a negligible scatter ($\Delta C < 0.3$ and $\Delta \epsilon < 0.2$) relative to the true input parameters. Filled red circles  indicate grid points in which a similarly large fraction of model galaxies simulated with these parameters is recovered within the original broad bin of concentration and ellipticity, but with a large scatter in concentration index ($\Delta C > 0.3$) and in ellipticity ($\Delta \epsilon > 0.2$) relative to the true input parameters. Squares identify  grid points, in a given bin of {\it observed} ellipticity and concentration,  that are dominated by a fraction of at least 50\% (yellow) to $>90\%$ (red)  of model galaxies which originate in a different bin of concentration or ellipticity, and are  scattered into  the observed grid point  by measurement errors\footnote{Note that red/green circles can be surrounded by squares, indicating that galaxies  born  with the given grid-point parameters are well-recovered within the same ellipticity/concentration bin, but, to this  indigenous  population, a population of  interloper  galaxies is added, due to scattering of galaxies into the given grid points from other ellipticity/concentration bins.}.  Finally, arrows indicate grid-points from which galaxies that are born with those given ellipticity/concentration parameters 'disappear' from that grid-point, and are observed with substantially different ellipticity and/or concentration parameters. Specifically, solid arrows indicate points of  the size-magnitude grid  in a  given {\it input/true} bin of ellipticity and concentration, from which at least $50\%$ of model galaxies  born  in this bin are scattered out from it, and end up, due to systematic measurement errors,  being 'observed' in other (incorrect) bins of ellipticity and concentration. Colors from yellow to red indicate an increasing fraction, above the threshold of 50\% required to plot an arrow (yellow), of out-scattered model galaxies (with red showing a loss of $>90\%$ of galaxies from the grid-point).  Empty arrows indicate grid-points from which similar fractions of galaxies ($>50\%$, yellow, up to $>90\%$, red) which are born with those given ellipticity/concentration values remain within the same ellipticity/concentration bin, but are recovered with errors in concentration $\Delta C> 0.3$ and ellipticity $\Delta \epsilon > 0.2$, respectively. As in Fig.\ \ref{f2}, the grey contours show the {\it corrected} $0.2 < z < 1.0$, $I_{814W} < 24$ mag COSMOS galaxy population(with a change in number density of a factor of two between density contours).  We discuss in Appendix \ref{testscorrs} some consistency checks that ensure that the derived correction functions  return quantities, including sizes, that are  free from  further biases.
 
We emphasize that  galaxies in some  regimes of parameter space can not be recovered, and thus some systematic biases can not be corrected. First, at very low surface brightnesses, the original Kron flux under-estimation causes objects to fall entirely below our $I_{814W} < 24$ mag catalog selection limit.  These regions of   classical   low surface brightness incompleteness  are identifiable in Figs.\ \ref{f2} and \ref{f3}.
This incompleteness effect has  a minimal impact on the observed magnitude-size distributions in COSMOS, hampering the observability of only the very largest galaxies with magnitudes fainter than $I_{814W} \sim 23.5$ mag, which motivates our initial sample selection criteria. Given its nature, this residual bias has no impact on the main results that we highlight in this paper.
Second, at very small sizes, due to PSF broadening, galaxies are  scattered  in regions of low concentration and low ellipticity values from bins of high concentration and ellipticity; there is thus an intrinsic degeneracy between small galaxies which are intrinsically diffuse and round, and galaxies which appear so although they are in reality more concentrated and elongated than they appear. These residual biases  must be kept in mind when analyzing galaxy samples;  again, they do not affect however  the results that we present in this paper.

Finally, in Appendix  \ref{galfitsizes} we show that corrections to  galactic sizes as a function of magnitude, size itself, concentration and ellipticity are  needed not only for aperture size measurements, but generally also for sizes that are  derived from analytical fits to the galaxy surface brightness distributions.  As a showcase we present in this Appendix the correction functions, in a similar way as in Figures \ref{f2} and \ref{f3}, but for sizes obtained using the public code Galfit \citep{pen02}. The Galfit analysis in the Appendix uses the same set of simulated galaxies which were used to correct the {\it ZEST+} size measurements  that we use in our analysis. We note that  the trends and  strengths of the correction vectors for the Galfit sizes  differ from  those reported in this section for the {\it ZEST+} sizes: Galfit does better than {\it ZEST+} in the regime of small sizes (i.e., smaller than the PSF), but is less accurate  in recovering sizes of large, low surface brightness galaxies.  Our main message here is twofold: first, in order to study galaxy populations spanning a large range of sizes and surface brightness, both aperture-based and the model-fit  size measurements require corrections. Second, once these corrections are applied, galaxy sizes derived with either of the two approaches are robust and in very good agreement with each other, as shown in Appendix \ref{galfitsizes}.  

We therefore stress that the results that we present in this paper  do not depend on our specific choice for how to quantify galaxy sizes.  We also expect our results to be unaffected by `morphological K-corrections', as remarked already in the Introduction,  due to the lack of  strong color gradients in high-z Q-ETG demonstrated in several earlier works (e.g.,  Toft et al.\ 2007; Guo et al.\ 2011).

\section{Evolution in  the Number Densities of Q-ETGs at Fixed Mass, Size and Surface Mass Density}\label{allresults}

\begin{figure*}
\epsscale{1.1}
\plotone{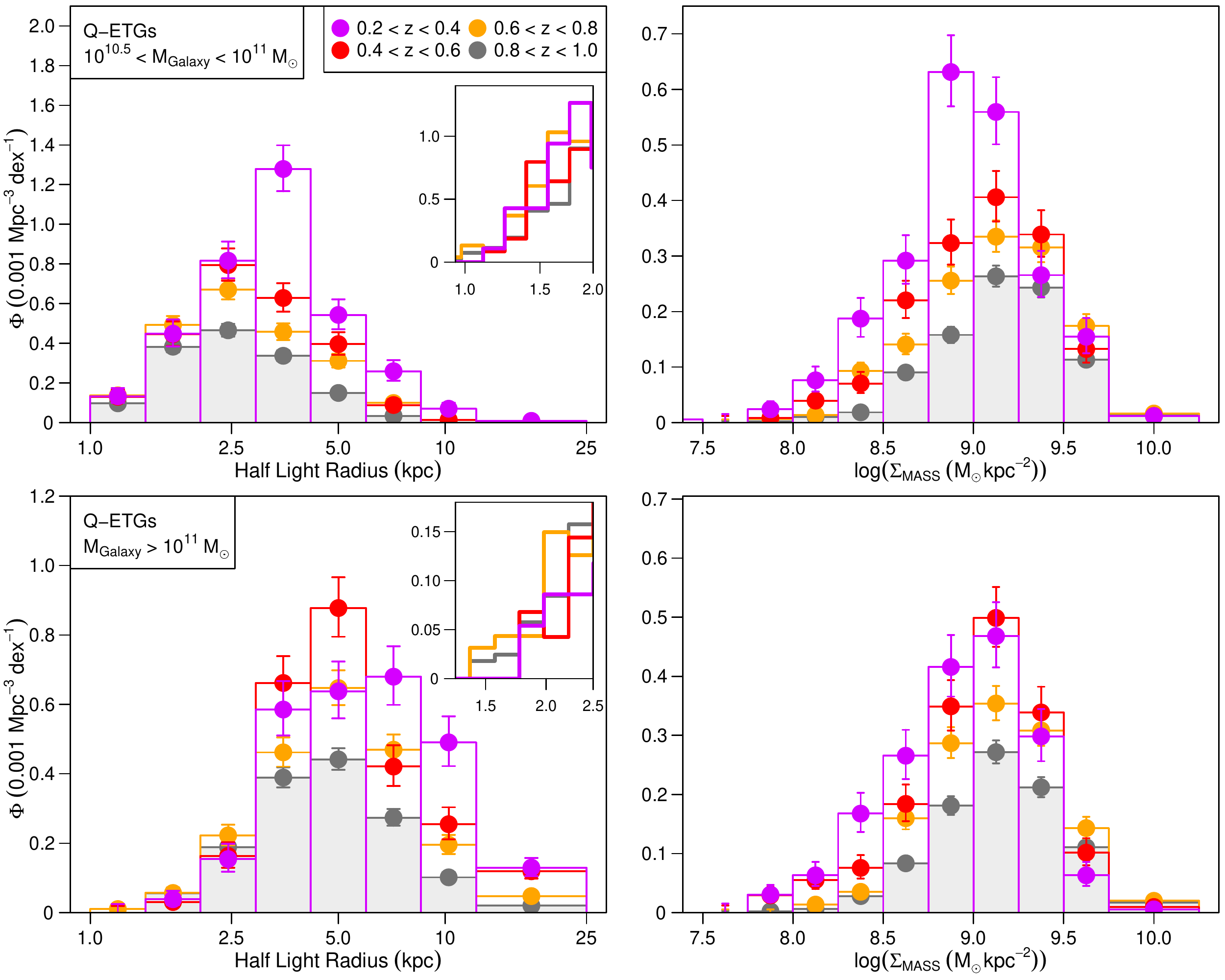}
\caption{The size (i.e., half-light radius) and surface mass density $\Sigma_\mathrm{MASS}$ distributions of quenched early-type, E/S0 galaxies (Q-ETGs) in COSMOS, in the redshift interval  $0.2 < z < 1$.  Results are shown for two bins of stellar mass, $10^{10.5} < M_\mathrm{Galaxy} < 10^{11}$$M_\odot$ (top row) and $M_\mathrm{Galaxy} > 10^{11}$$M_\odot$ (bottom row).  
The redshift interval is split in four bins, i.e., $0.2 < z < 0.4$ (magenta), $0.4 < z < 0.6$ (red), $0.6 < z < 0.8$ (orange), $0.8 < z < 1.0$ (grey). 
The error bars on each histogram represent only the Bayesian 1$\sigma$ CIs accounting for the Poisson noise in each bin.  As noted in Section \ref{dataset}, photometric sample incompleteness and the non-Poissonian component of cosmic variance contribute further uncertainties on the order of 15\% each to the absolute normalization.  Note also that the normalization of the $0.4 < z < 0.6$ and $0.6 < z < 0.8$ histograms have been adjusted against cosmic variance in COSMOS, assuming a   galaxy number density of galaxies with $M_\mathrm{Galaxy} > 10^{10.5}M_\odot$ that linearly increases with redshift, as constrained by the redshift bins $0.2 < z < 0.4$ and $0.8<z<1.0$ (i.e., excluding the redshift bins $0.4<z<0.6$ and $0.6<z<0.8$, which, in the COSMOS field,  are known to contain a severe underdensity and a massive cluster, respectively).  The insets in the size-function plots present a zoom on the $<2$ kpc ($<2.5$ kpc in the high mass bin) part of the histograms:  in the low mass bin, we do not observe any `disappearance'  towards lower redshifts of compact Q-ETGs  down to the smallest $\sim1$kpc sizes; at high masses,  a disappearance towards lower redshifts of compact galaxies is observed, which appears to be increasingly stronger the smaller the galaxy size. \label{f4}}
\end{figure*}

Fig.\ \ref{f4} shows the number density of Q-ETGs in COSMOS at fixed size (i.e., the  size-function  $\Phi_{r_{1/2}}$; lefthand panels) and surface mass density $\Sigma_\mathrm{MASS}$ (i.e, the  $\Sigma_\mathrm{MASS}$-function  $\Phi_{\Sigma_\mathrm{MASS}}$; righthand panels) in four bins of redshift: $0.2 < z < 0.4$ (magenta), $0.4 < z < 0.6$ (red), $0.6 < z < 0.8$ (orange), and $0.8 < z < 1.0$ (grey).  

We restrict our investigation  to two fixed intervals of stellar mass above the effective completeness limit at $z=1$ for the  spectro-morphological 'early-type galaxy' class in the $I_{814W}<24$ COSMOS catalog (see \citealt{oes10,ilb10}). In particular, the large size of the  COSMOS sample enables us to split the total sample of massive, $>10^{10.5} M_\odot$ Q-ETGs at these redshifts in two distinct mass bins straddling across the   $M_{Galaxy} =  10^{11} M_\odot \sim M^*$ value. We thus define  a 'low mass bin', with $10^{10.5}  < M_\mathrm{Galaxy} < 10^{11} $$M_\odot$ (top row in Fig.\ \ref{f4}), and a 'high mass bin', with $M_\mathrm{Galaxy} > 10^{11}$$M_\odot$ (bottom row).   We have chosen  a linear scale on the y-axis to display the size and $\Sigma_\mathrm{MASS}$ functions $\Phi$; the error bars, however, reflect the Bayesian 1$\sigma$ confidence intervals (CIs) on the number density in each bin treated as a Poisson rate parameter (obtained from inversion of  the posterior probability distribution for $\Phi$,  given the observed number of objects in that bin and the improper uniform prior; see \citealt{kra91} for details). 

Note that, in order to correct for cosmic variance effects in the COSMOS field, well known to contain  a severe under-density  and a massive cluster at $z\sim0.5$ and $z\sim0.75$, respectively (see, e.g.\ \citealt{sca07b,oes10}), we have followed the approach of  \cite{oes10} and normalized the total number density  $\Phi (M_\mathrm{Galaxy} > 10^{10.5} M_\odot$)  of  galaxies of all morphological types in the $0.4 < z < 0.6$  and $0.6 < z < 0.8$ redshift bins, and with stellar masses $M_\mathrm{Galaxy}> 10^{10.5}M_\odot$, against a linear interpolation with redshift of the  number densities of corresponding galaxies in the  $0.2<z<0.4$  and $0.8 < z < 1.0$ redshift bins.   

\begin{figure*}
\epsscale{1.1}
\plotone{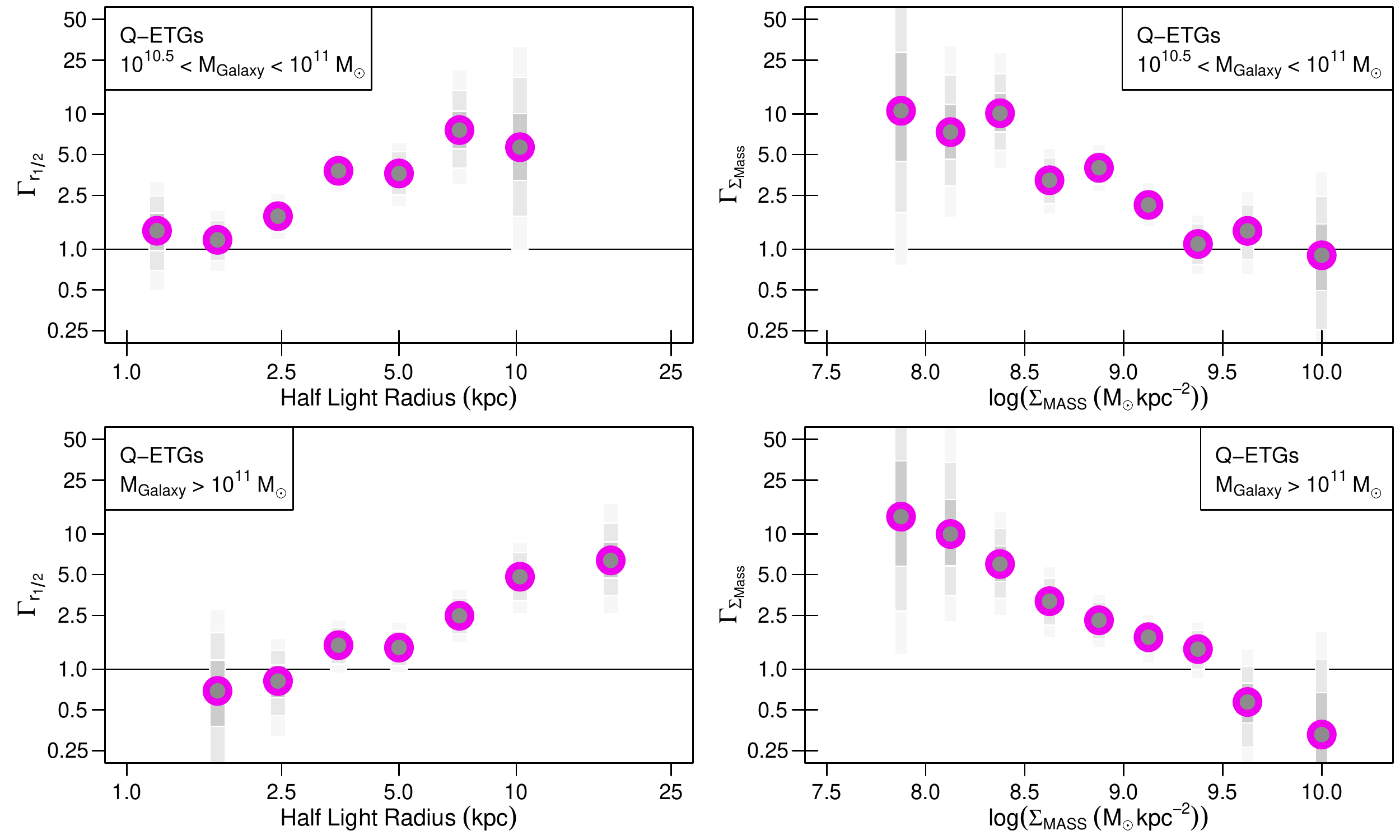}
\caption{The growth  factors between $z=1$ and $z=0.2$, $\Gamma_{r_{1/2}}$ and $\Gamma_{\Sigma_\mathrm{MASS}}$, of the number density  of quenched early-type galaxies (Q-ETGs)  at fixed size (left) and  $\Sigma_\mathrm{MASS}$ (right). Results are shown separately for the low mass bin, $10^{10.5} < M_\mathrm{Galaxy} < 10^{11} M_\odot$ (top row) and for the high mass bin, $M_\mathrm{Galaxy} > 10^{11} M_\odot$ (bottom row).  The growth factors are defined as the ratios between the size-functions (left) and $\Sigma_\mathrm{MASS}$-functions (right) at $0.2 < z < 0.4$,  and the corresponding  functions at $0.8< z <1$.  The dark-, intermediate-, and light-grey vertical bars trace the 1$\sigma$, 2$\sigma$, and 3$\sigma$ Bayesian confidence intervals on these factors, respectively, with the medians marked by the magenta+grey datapoints. The coloring of these data points is used as a reminder that they express the ratios between the magenta and grey curves in Fig.\ \ref{f4}. \label{f5}}
\end{figure*}

To better highlight the global evolution since $z=1$ of the size- and $\Sigma_\mathrm{MASS}$-functions, we plot in Fig.\ \ref{f5} the ratios between these functions in the lowest ($0.2 < z < 0.4$)  and highest ($0.8 < z <1$) redshift bins of our analysis.  These ratios, defined respectively as:

\begin{equation}
\Gamma_{r_{1/2}}=\frac{\Phi_{r_{1/2}}(0.2<z<0.4) }{ \Phi_{r_{1/2}}(0.8<z<1)}
\end{equation}
 
 and 
 
 \begin{equation}
\Gamma_{\Sigma_\mathrm{MASS}}=\frac{\Phi_{\Sigma_\mathrm{MASS}}(0.2<z<0.4) }{ \Phi_{\Sigma_\mathrm{MASS}}(0.8<z<1)}, 
\end{equation}

highlight in a straightforward manner the rate of growth  in the  number densities of Q-ETGs  in each bin of size and 
$\Sigma_\mathrm{MASS}$. 

At the highest stellar masses, i.e., in the $M_\mathrm{Galaxy} >10^{11} M_\odot$ bin of our analysis,
we detect a decrease  from  $z=1$ to $z=0.2$ of about  30\%-40\% of the smallest and densest  Q-ETGs, i.e.,  of high-mass Q-ETGs with half-light radii smaller than $\sim2$ kpc.    Note that in our data this is only seen at the  $1-\sigma$-level. Nevertheless this result,  obtained within a self-consistent dataset, formally agrees with the several works which report a decreasing number density of  compact  galaxies with increasing age of the Universe (see references above).    At these high masses,  the disappearance towards lower redshifts of compact galaxies   seems to be increasingly larger the smaller the galaxy size, as shown in the inset  in the relevant size-function plot of Figure \ref{f4}, which   zooms on the $<2.5$ kpc scales.   

The analysis of our  low-mass  bin at $10^{10.5} < M_\mathrm{Galaxy}< 10^{11} M_\odot$, i.e., just below $M^*$,  also shows surprises. The number density of the  $r_{1/2}\leq 2$ kpc  Q-ETGs   remains remarkably stable throughout the $z = 1 \rightarrow 0.2$  redshift range. At these masses, there is essentially no change in the number density of the most compact Q-ETGs  since $z\sim1$ down to $z\sim0.2$, and thus  no evidence for any disappearance of  compact galaxies due to an increase in their individual sizes  across this time span.   This results is not affected by our binning in size: it holds  down to the smallest $\sim1$kpc sizes, as shown in the  inset in the relevant size-function panel of Figure \ref{f4}, which plots again a zoom-in of  the small-scale size-function (this time in the $<2$kpc regime). 
Note that we ascertain this constancy self-consistently within the COSMOS sample, without recurring to comparisons between different studies or datasets. This ensures no  spurious effects introduced by different approaches for computing the sizes, and by different PSFs, noise properties, systematic effects, different depths, and other such potential complications.
 
Similarly striking in Fig. \ref{f5} is that, in both mass bins,  we do observe a  strong evolution in the size and $\Sigma_\mathrm{MASS}$ functions of Q-ETGs, with a marked dependence on galaxy size in both mass bins for the population growth rates  $\Gamma_{r_{1/2}}$. This evolution involves however not the small-size,   compact  Q-ETG population, but rather the number densities of {\it large-size} (and thus {\it low $\Sigma_\mathrm{MASS}$}) Q-ETGs. In particular, the inspection of Fig.\ \ref{f5}  shows that,  in both  low and high mass bins, a substantial increase in number density of large-size Q-ETGs has taken place since $z=1$:  the  growth rates  $\Gamma_{r_{1/2}}$ rise, respectively in the low mass and high mass bins, from of order unity (i.e., the null growth) and from the  negative rate  of $\sim 0.3$ at $r_{1/2} \lesssim 2$ kpc, as reported above, 
 up to $\sim$5-6 by $r_{1/2} \gtrsim 5$kpc (low mass bin) and $r_{1/2} \sim 12$ kpc (high mass bin).
This corresponds to similar growth factors from $z=1$ to $z=0.2$ for the population of {\it low-$\Sigma_\mathrm{MASS}$}   galaxies, i.e.,  growth rates $\Gamma_{\Sigma_\mathrm{MASS}}\sim 5$  from $z=1$ to $z=0.2$  of galaxies with  $\Sigma_\mathrm{MASS} \leq 9$ $M_\odot$ kpc$^{-2}$.

\begin{figure*}
\epsscale{1.1}
\plotone{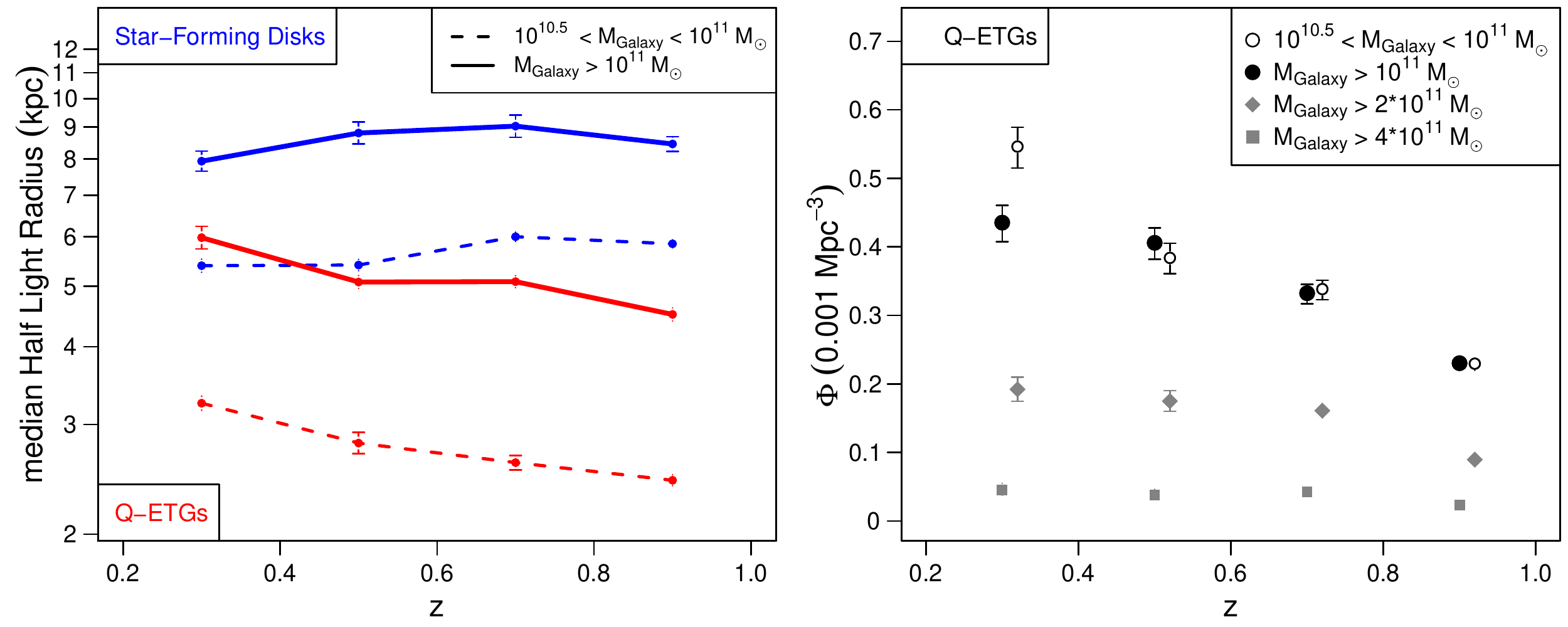}
\caption{{\it Left panel:} The median size (half light radius) of Q-ETGs (in red) and star-forming disks (in blue),  as a function of redshift, for the high mass bin ($M_\mathrm{Galaxy}> 10^{11}$ $M_\odot$; solid lines) and low mass bin ($10^{10.5}$ $M_\odot<M< 10^{11} M_\odot$; dashed lines).  {\it Right panel:}   The number density of Q-ETGs  as a function of redshift in the low- (empty circles) and  high-mass  (solid circles) bin. Error bars indicate $1\sigma$ uncertainties on the median values. We also plot the number density evolution of $2\times 10^{11} M_\odot$ (diamonds) and $4\times 10^{11} M_\odot$ (squares) to show that most of the growth in our high mass bin occurs peaked at $\sim 10^{11} M_\odot \sim M^*$. \label{f6}}
\end{figure*}

The significant increase in number density of large-size  Q-ETGs with cosmic time between $z\sim1$ and $z\sim0.2$ implies that the {\it newly-quenched, large  Q-ETGs} (hereafter  NQ-ETGs\footnote{ Note that  the acronym ``NQ-ETGs'' is used to indicate ETGs that are newly quenched at any given epoch, 
not an observational  galaxy sample selected to span a specific range of values in  some measured parameters. In Figure \ref{f8}, the population of NQ-ETGs which emerges between our highest and lowest redshift bin is shown, in each of the mass bins, by the black hatched histogram that results from  subtracting  the $ 0.8<z<1$ size function  from the $0.2<z<0.4$ size function.}) lead to  a substantial increase in the {\it median} half-light size for the whole Q-ETG population.
The proportional increase in median size  is comparable in the high mass  and  low mass bins, and equal to  a factor of $\sim$1.3 over the redshift span of our analysis. Specifically, as shown in Fig.\ \ref{f6} (left panel), the median size of Q-ETGs  grows from $4.5\pm 0.1$ kpc to 6.0$\pm 0.3$ kpc  in the high mass bin (solid red line), and from $2.4\pm 0.1$ kpc to 3.2$\pm 0.1$ kpc in the low mass bin (dashed red line), respectively; these imply a formal growth as approximately $(1+z)^{-0.70\pm{0.15}}$.
 Correspondingly, and consistently with other analyses, there is  an increase by a factor of $\sim2-2.5$, between $z\sim1$ and $z\sim0.2$, in the overall number density of $M_\mathrm{Galaxy} > 10^{10.5} M_\odot$  Q-ETGs. This is shown  in the right panel of Fig.\ \ref{f6}, where we plot the number density of Q-ETGs, summed over all sizes,  as a function of redshift; the result is shown separately for our low-mass ($10^{10.5} M_\odot < M_\mathrm{Galaxy} < 10^{11} M_\odot$; empty points) and high-mass ($M_\mathrm{Galaxy} > 10^{11} M_\odot$; solid points) bins.  As discussed in Section \ref{intro}, these  number density growth factors for Q-ETGs agree well with the measured evolution of the mass function of quiescent galaxies since $z\sim1$.

Note that in Figure \ref{f6}  the increase since $z=1$  of the number density of Q-ETGs with masses integrated above $10^{11} M_\odot$   contrasts the well-known constancy in number density of the most massive Q-ETGs, with masses of order $2-4\times 10^{11} M_\odot$ (see also  Cimatti et al.\ 2006; Scarlata et al.\ 2007b; Pozzetti et al. 2010; Ilbert  et al.\ 2013). This highlights that  the number growth from $z=1$ to $z=0$ of the  Q-ETG population   at galaxy masses above $10^{11} M_\odot$ occurs predominantly within a very narrow mass peak around $\sim 10^{11} M_\odot \sim M^*$. This reinforces the notion that the probability for a galaxy to survive quenching and keep living as a star-forming galaxy    drops exponentially with increasing stellar mass and becomes negligible above $M^*$ (see Peng et al.\ 2010).

Keeping in mind that we observe little or no decrease in the number densities of the smallest and most compact Q-ETGs since $z=1$, the key question to answer is  what is the main cause for the  appearance of  large  Q-ETGs at later times -- appearance which is  largely responsible for the measured increase in the median size of the whole population of such quenched systems with an early-type  morphology, whereas the growth of individual galaxies has apparently no effect on it.

\section{Discussion: The Emergence of Large Q-ETGs at Late Epochs}\label{discussion}

We can summarize our findings above with two simple statements: $(i)$ the number density of compact $<2$ kpc Q-ETG remains stable since $z=1$, with at most a very modest decrease with decreasing redshift of order 30\% for the most massive $>10^{11} M_\odot$ Q-ETGs; and $(ii)$ most of the growth of the median size of the $>10^{10.5} M_\odot$ Q-ETG population is due to the appearance at later times of large(r) Q-ETGs.

On the basis of the current evidence, it is possible of course that the Q-ETG population  grows in size with cosmic time in a coordinated manner, so as to keep constant the number density of the smallest/densest galaxies, which should be created   at the same rate at which  they would be shifted to the higher size bins. While it seems  somewhat contrived that such  conspiracy  may take place, we note  that   in this case the most compact galaxies would be the last that have been quenched, and thus they should be systematically younger than average. Current analyses of the stellar population ages in the $z\sim2$ \citep{saracco} and $z=0$ \citep{vanderWel} Q-ETG populations  find hints for an effect in the opposite direction, i.e., the most compact galaxies look older, not younger, than their larger counterparts.

\subsection{The rest-frame $(U-V)$ colors of compact and large Q-ETGs}

\begin{figure*}
 \epsscale{1.1}
\plotone{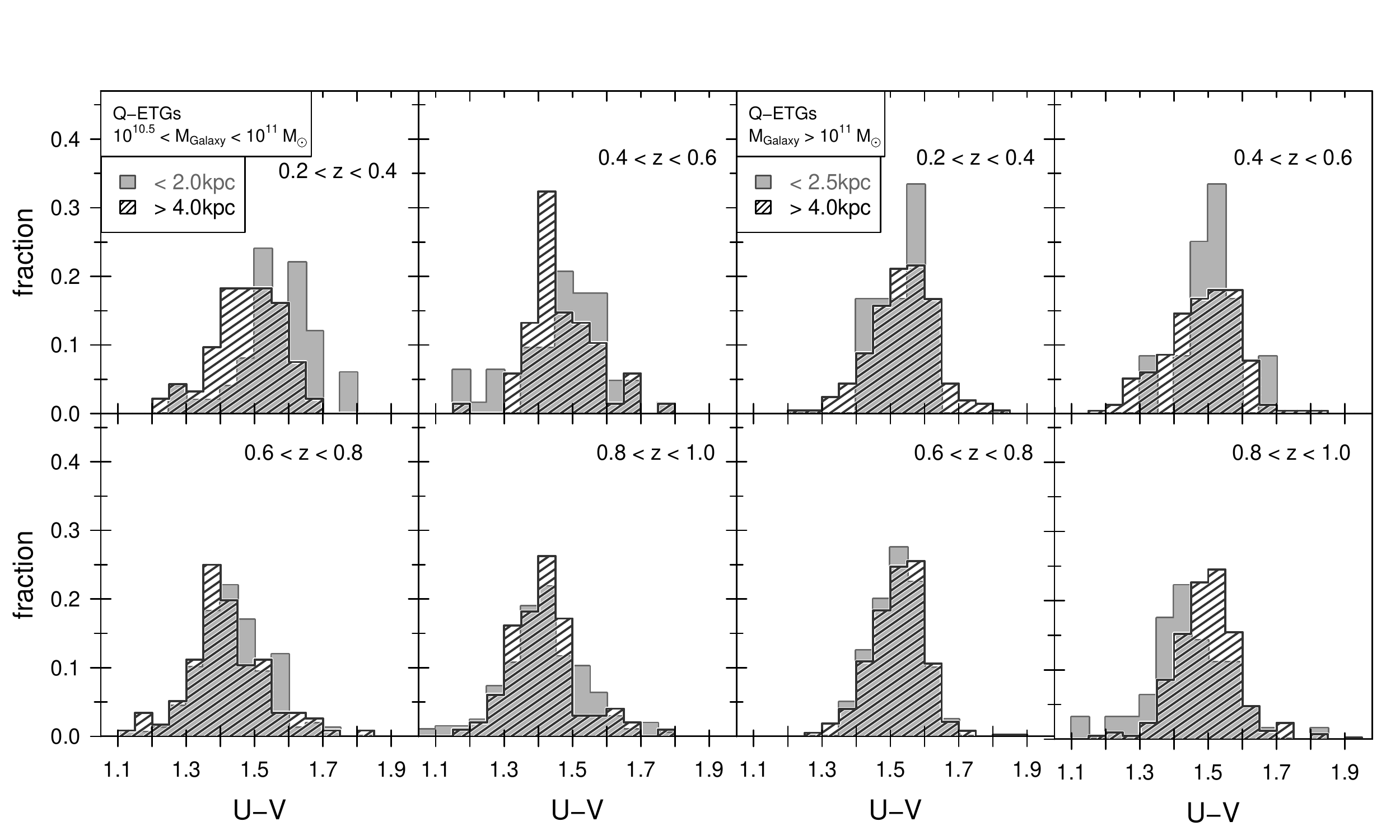}
\plotone{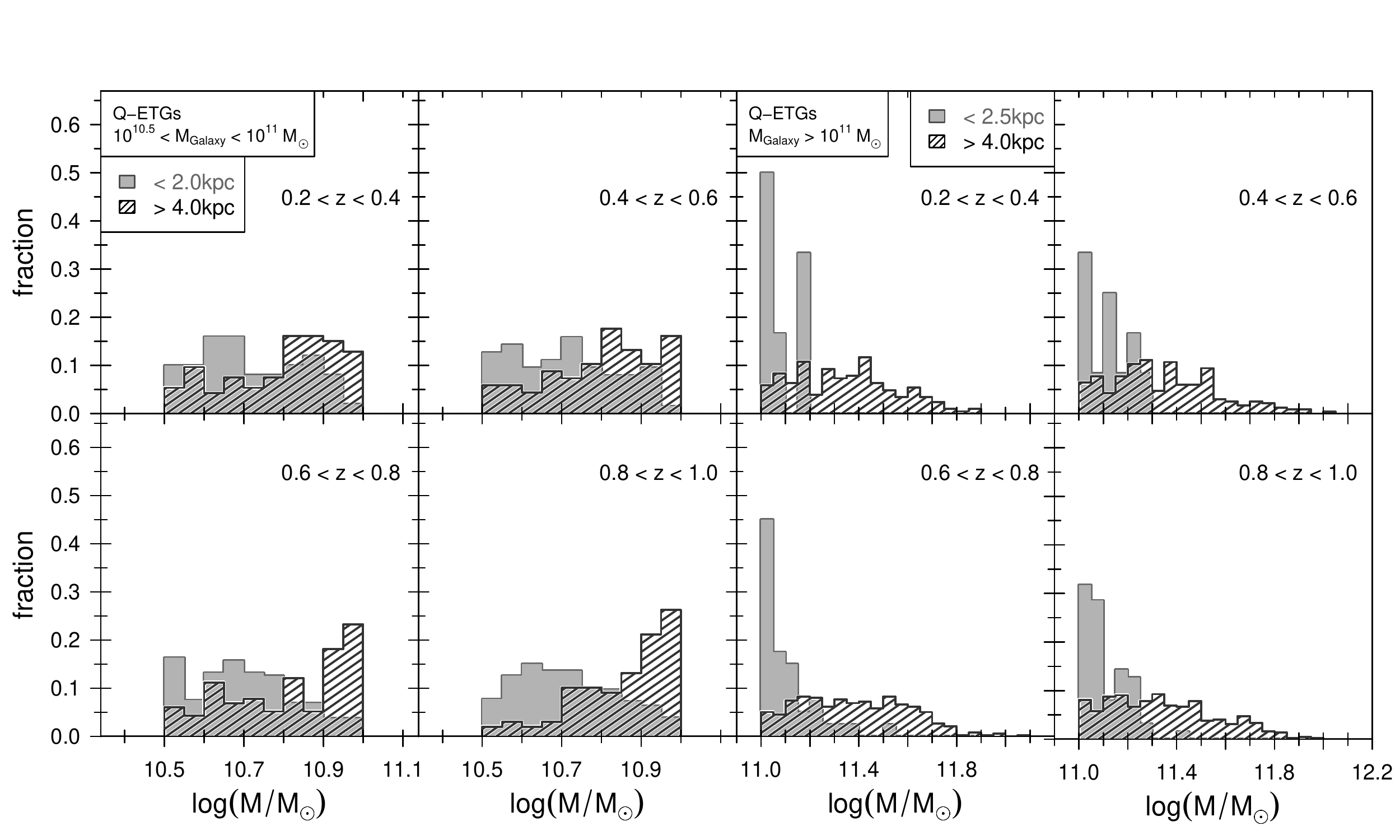}
\caption{{\it Top:} The rest-frame $(U-V)$ color  distributions of Q-ETGs in COSMOS, in the redshift interval  $0.2 < z < 1$.  Results are shown for two bins of stellar mass, $10^{10.5} < M_\mathrm{Galaxy} < 10^{11}$$M_\odot$ (two leftmost columns) and $M_\mathrm{Galaxy} > 10^{11}$$M_\odot$ (two rightmost column).  
The samples are split in the same four redshift intervals of Fig.\ \ref{f4}. Each panel compares the distributions of colors for Q-ETGs larger than 4 kpc and smaller than 2 kpc (or 2.5 kpc, in the low and high mass bin, respectively).  {\it Bottom:} The corresponding stellar mass distributions for the compact and large samples of Q-ETGs of the top figure. 
\label{f7}}
\end{figure*}

We search for possible  trends  in stellar population ages of compact and large Q-ETGs in our own data: in Fig.\ \ref{f7} (top), we plot the distributions of the rest-frame $(U-V)$ colors of the low (two leftmost columns) and high (two rightmost columns) mass bins, split in two subsamples with sizes, respectively, $<2$ kpc (for the low mass galaxies, and $<2.5$ kpc, for the high mass galaxies)
and $>4$ kpc. The median colors of both populations in each redshift bin are listed in Table \ref{ColorTable}. 

It is immediately evident that there is a trend in the same direction as the previous studies at lower and higher redshifts. The colors of the most compact  Q-ETGs become  on average increasingly redder towards lower redshifts. Specifically, the color difference between compact Q-ETGs at $z=1$ and at $z=0.2$ is  $\Delta(U-V)\sim0.16$, which is in very good agreement with the expected
color change  for a single stellar population that formed at $z\sim2$
and passively evolves between these two epochs. Furthermore, at low masses, compact Q-ETGs
are systematically redder than  their large counterparts at similar masses, strengthening the interpretation that the former are older than the latter (see also Shankar 
\& Bernardi 2009; Saracco et al.\ 2011).

\begin{deluxetable*}{lrrrr}
\tablecolumns{5}
\tablewidth{0pt}
\tablecaption{Median rest-frame $(U-V)$ colors of compact   (low mass bin: $<2$ kpc; high mass bin: $<2.5$ kpc) and  large ($>4$ kpc) Q-ETGs}
\tablehead{\colhead{redshift} & \multicolumn{2}{c}{$10^{10.5}<M/M_\odot<10^{11}$} &  \multicolumn{2}{c}{$M/M_\odot>10^{11}$}\nl\nl
\colhead{} &\colhead{$r_{1/2}<$ 2 kpc} &\colhead{$r_{1/2}>$ 4 kpc} &\colhead{$r_{1/2}<$ 2.5 kpc}&\colhead{$r_{1/2}>$4 kpc} }
\startdata
$0.2 \le z <0.4$ &$1.57\pm0.01$ & $1.49\pm0.01$ & $1.56\pm0.04$ & $1.55\pm0.01$\nl  
$0.4 \le  z <0.6$ &$1.49\pm0.01$ & $1.45\pm0.01$ & $1.52\pm0.04$ & $1.48\pm0.01$\nl  
$0.6 \le  z <0.8$ &$1.44\pm0.01$ & $1.41\pm0.01$ & $1.53\pm0.02$ & $1.53\pm0.01$\nl  
$0.8 \le   z <1.0$ &$1.42\pm0.01$ & $1.41\pm0.01$ & $1.42\pm0.02$ & $1.50\pm0.01$ \nl
\enddata
\label{ColorTable}
\end{deluxetable*}

\begin{deluxetable*}{lrrrr}
\tablecolumns{5}
\tablewidth{0pt}
\tablecaption{Median stellar masses within the corresponding mass bins  for the samples of Table \ref{ColorTable} of compact   and  large Q-ETGs }
\tablehead{\colhead{} & \multicolumn{2}{c}{$10^{10.5}<M/M_\odot<10^{11}$} &  \multicolumn{2}{c}{$M/M_\odot>10^{11}$} \nl \nl
\colhead{redshift} &\colhead{$r_{1/2} < $ 2 kpc} &\colhead{$r_{1/2} > $ 4 kpc} &\colhead{$r_{1/2} < $ 2.5 kpc}&\colhead{$r_{1/2} > $ 4 kpc} }
\startdata
$0.2 \le  z < 0.4$ &$10.69\pm0.03$ & $10.83\pm0.02$ & $11.06\pm0.04$ & $11.34\pm0.02$\nl  
$0.4 \le  z < 0.6$ &$10.71\pm0.02 $ & $10.82\pm0.02$ & $11.11\pm0.04$ & $11.32\pm0.02$\nl  
$0.6 \le  z < 0.8$ &$ 10.69\pm0.01$ & $10.83\pm0.03$ & $11.07\pm0.02$ & $11.36\pm0.02$\nl  
$0.8 \le  z < 1.0$ &$10.70\pm0.01$ & $10.89\pm0.02$& $11.07\pm0.01$ &  $11.32\pm0.01$  \nl
\enddata
\label{MassTable}
\end{deluxetable*}

We test that variations in stellar mass within the formal mass bins do not impact the conclusions above concerning the relative average ages for the compact and large Q-ETG populations. Also in Fig.\ \ref{f7} (bottom) we show the distributions of stellar masses, within each redshift bin and formal bin  of stellar mass, for the same compact and large samples of Q-ETGs of the color analysis. The median stellar masses of these distributions are listed in Table \ref{MassTable}.  In each formal bin of mass,  the median stellar mass of the compact population is very constant with redshift. There is understandably a trend for the compact Q-ETG sample to have, within each formal mass bin, a slightly smaller median mass than the large Q-ETG population.   This goes however in the direction of decreasing, if anything,  the color difference between old  compact versus young  large Q-ETGs. We thus conclude that the color difference   that we have detected   between compact and large Q-ETGs in the low mass bin is a genuine stellar population effect, indeed consistent with a younger average age of   large relative to compact Q-ETGs.  In the high mass bin the two populations, compact and large, have much more similar colors. At these high masses mergers play a much larger role. We will further comment on this below.
  
In the light of these results, a   more plausible interpretation is that a  static  rather than  dynamic  equilibrium holds for the number density of   $<2$ kpc   Q-ETGs (i.e., the population of compact Q-ETGs remains virtually unchanged between $z=1$ and $z=0.2$, without either creation of new compact Q-ETGs or growth of their individual sizes over this time period).  The emergence at low redshifts of the population of large  and   diffuse  Q-ETGs, seen in both our low- and high-mass bins,  therefore results from larger  star-forming  galaxies experiencing the  quenching trauma  at later times.

\subsection{Comparison of typical sizes of newly-quenched galaxies at different redshifts}

We can use the size functions of Q-ETGs of Fig.\ \ref{f4} to derive, in both our mass bins,  empirical  estimates for the size functions of the newly-quenched populations over the $z = 1 \rightarrow 0.2$ time period, and their median sizes at the median redshift of $z \sim 0.6$.  These estimates are readily  defined through  the {\it difference} between the $0.2 < z < 0.4$  and the $0.8 < z < 1.0$ size-functions of Q-ETGs of Fig.\ \ref{f4}, replotted in Fig.\ \ref{f8} as solid magenta and grey histograms, respectively   (left and right panels refer respectively to the low and high mass bins). The size functions of the NQ-ETGs are shown as a black hatched histograms; the corresponding median half-light radii of  the  NQ-ETGs are shown as black arrows in Fig.\ \ref{f8}. These median sizes are equal to 4.3$\pm$0.4 kpc and 8.7$\pm$0.9 kpc in the low and high mass bins, respectively.

Franx et al.\ (2008) report a median size of $\sim2$ kpc for  Q-ETGs at $z = 2.0$  in a mass bin equivalent to our lower mass bin.  These high-redshift Q-ETGs are all likely to have been recently quenched at the time that we observe them.  Within the large uncertainties involved, the change  in size of the newly-quenched galaxies is therefore consistent with the idea that NQ-ETGs have, at fixed mass, a size that scales roughly as $(1+z)^{-1}$, i.e. with the cosmic scaling of the sizes of dark matter halos of a given mass.  Equivalently, the mean stellar density of NQ-ETGs appears to scale roughly as the mean density of the Universe.  

This idea may then naturally explain the strong connection between sSFR and the mean stellar density of galaxies in the local Universe.   Quenched galaxies with low sSFR would naturally have higher stellar densities simply because they would, generally, have been quenched at significantly earlier epochs.

\subsection{Consistency with predictions from a continuity equation}\label{conteq}

We finally check whether the global  number density growth factors that we have observed for the Q-ETGs between $z=1$ and $z=0.2$ are also roughly in agreement with those expected by applying a   number continuity equation  to the time-evolution of the star-forming galaxy population, assuming that this is quenched without a substantial increase of stellar mass in the post-quenching phase (Peng et al.\ 2010).  Anchoring their analysis to SDSS \citep{yor07} and zCOSMOS \citep{lil07,lil09} data, these authors  show strong evidence that,   while 'environment-quenching' dominates at lower masses,  the dominant process that piles up galaxies at  $>10^{10.5} M_\odot$  onto  the red-and-dead sequence  is
a quenching mechanism  linked only to  mass  and not to environment (i.e., their 'mass quenching';
 cf. Figure 13 in Peng et al. 2010).  
 
To  explore whether mass-quenching of star-forming   galaxies is quantitatively consistent, in terms of numbers and sizes, with the evolution of the size-function of Q-ETGs that we have observed, $(i)$ we  compute, in our two mass bins, the size (and $\Sigma_\mathrm{MASS}$) functions  for the {\it star-forming  population} in the COSMOS dataset; and $(ii)$ we adopt the predictions of Peng et al.\ (2010) for 
sudden environment-independent  mass-quenching    of  star-forming galaxies,  which is the relevant quenching mode  at the mass scales of our  sample.  

We choose to limit this analysis to the star-forming {\it disk} galaxies in our COSMOS sample. We note that, at the stellar masses that we are investigating, disks represent $>80\%$ of the star-forming population; furthermore, we also remark that, above $10^{10.5} M_\odot$, the population of star-forming disk  galaxies is largely, $>90\%$,  made of systems which have a substantial bulge component. In quantitative comparisons with previous studies, it is thus important to keep in mind that  the  star-forming population that we consider below is essentially a  sample of   star-forming disks with a large if not dominant bulge component.
We stress again however that we have tested that our results remain unchanged if the morphological selections are removed from our study. Apart from relatively small differences in the normalizations of the quenched and star-forming populations, which largely compensate each other, the only marginally detected difference is a  higher number of  large-size, {\it irregular} star-forming galaxies in the high relative to low redshift bin. This is not unexpected, and does not alter any of our conclusions. Our choice to select a morphologically well-behaved samples of star-forming disks  ensures   a higher reliability in the estimates of galactic sizes relative to  disturbed, irregular morphologies.

Fig.\ \ref{f9} shows the size- and $\Sigma_\mathrm{MASS}$ functions of star-forming  disk galaxies in the same redshift and mass bins as in Fig.\ \ref{f4}. Again we have chosen here a linear scale for $\Phi$ on the vertical axis, and  error bars reflecting the Bayesian 1$\sigma$ CIs on the number density in each bin treated as a Poisson rate parameter.  Note that star-forming galaxies do not disappear upon quenching, as they are continuously
replenished from lower mass bins, thanks to star formation. Actually, they keep increasing their number density (except at the smallest sizes), as indeed found for their mass function both in Ilbert et al.\ (2010, 2013) and  in the Peng et al.\ (2010) model. 

The  median half-light sizes of the low mass (dashed curve) and high mass (solid curve) star-forming  population as a function of redshift are  furthermore plotted as blue lines in the left panel of Fig.\ \ref{f6}. 
Note that the star-forming disk galaxy population  has typically an average  median size ($\Sigma_\mathrm{MASS}$) a factor of  $\sim1.5-2$ larger (lower) than that of Q-ETGs of similar mass and redshift. Also, the median size  ($\Sigma_\mathrm{MASS}$)  of star-forming disk galaxies  shows, within the errors, a remarkable constancy with redshift over the   $z=1$ to $z=0.2$ period.  These results are in qualitative agreement with other work showing that the median sizes of star-forming galaxies are larger and evolve much slower than those  of quenched galaxies (e.g., Franx et al.\ 2008). Quantitatively, however, these and other authors suggest a stronger    evolution of the average size of star-forming galaxies than what we find. It is difficult to compare at these later redshifts results  based on different samples: stellar masses may be systematically different, the nominal mass bins may differ from ours, and   samples typically  heterogeneous, i.e.,  they use different galaxy samples to compare different redshifts. For example, Franx et al.\ use, as several other authors, the SDSS sample to set the $z=0$ reference, with few data points   at comparably high masses between $z=1$ and $z=0$. We thus refrain from attempting a direct comparison between our star-forming sample and others, and stress instead that,  using the self-consistent COSMOS sample  with  uniformly derived sizes and well-defined morphological properties as described above, we  actually detect a lack of evolution in the median size of $10>10^{10.5} M_\odot$ star-forming  (mostly bulge-dominated)  disk galaxies over the $z =1 \rightarrow 0.2$ redshift period. We defer to a follow-up analysis  to understand the impact of cosmological disk fading and other factors on such apparent lack of evolution in the median size of  disk star-forming galaxies in our sample. 

We then use Eqns.\ (1) and (27) of Peng et al.\ (2010) to estimate the fractions of star-forming galaxies that are predicted to  undergo mass-quenching  over the $z=1\rightarrow0.4$.  Details of the derivation are described  in Appendix \ref{peng}. For the lower mass bin, $10^{10.5} < M_\mathrm{Galaxy} < 10^{11} M_\odot$, this yields quenching fractions $f_{MQ}$ of star-forming galaxies  in the $0.8 < z < 1.0$, $0.6 < z < 0.8$ and  $0.4 < z < 0.6$ redshift bins of  0.22, 0.16 and 0.13, respectively.   At $M_\mathrm{Galaxy}>10^{11} M_\odot$, the fractions of newly-quenched galaxies  in the  same redshift bins are instead $f_{MQ}$ = 0.57, 0.44 and 0.35, respectively.  
 
\begin{figure*}[t!]
\epsscale{1.1}
\plotone{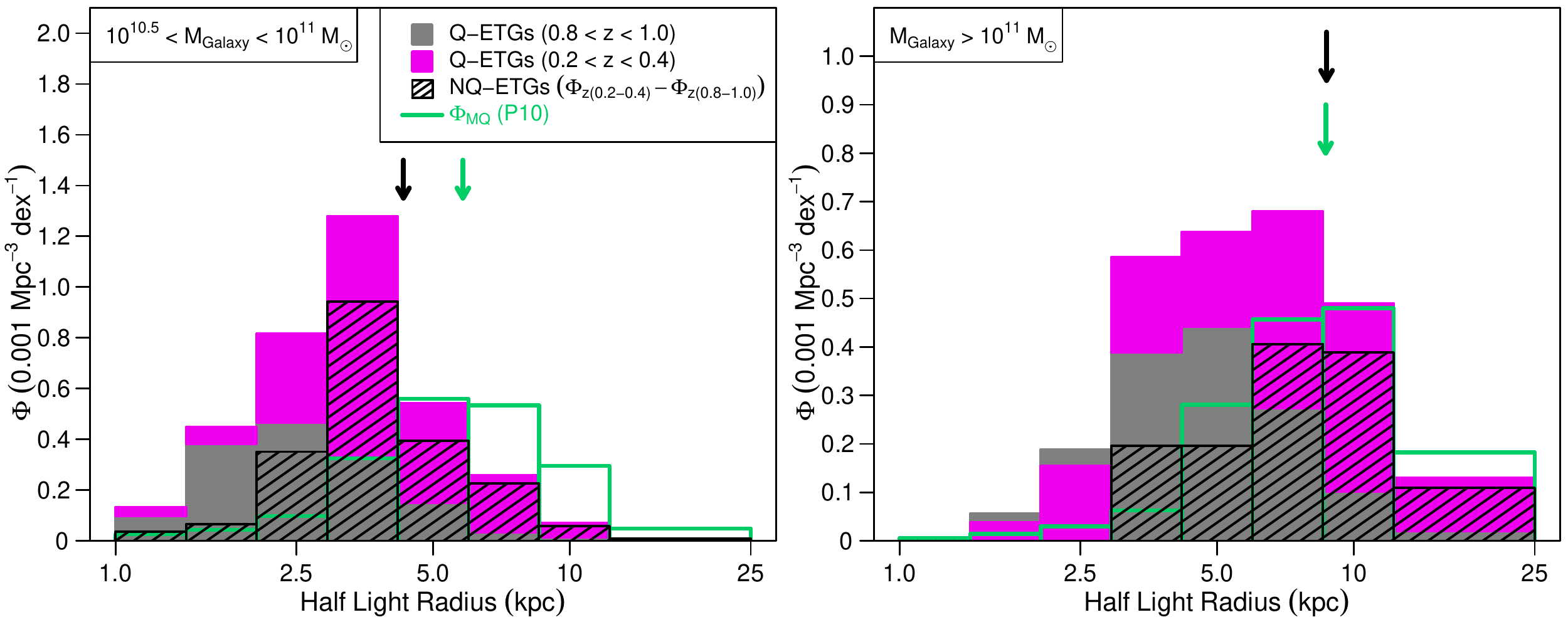}
\caption{ Empirical reconstruction for the size functions of  NQ-ETGs over the $z\sim1$ and $z\sim0.2$ time period, defined through  the {\it difference} between the $0.2 < z < 0.4$ (magenta histograms) and the $0.8 < z < 1.0$ (grey histograms) size-functions of Q-ETGs of Fig.\ \ref{f4}. Top and bottom rows are respectively for  $10^{10.5} < M < 10^{11}$ $M_\odot$  and $M_\mathrm{Galaxy}>10^{11} M_\odot$ galaxies. 
The black arrows show the median sizes of  these populations of NQ-ETGs over  the $z = 1 \rightarrow 0.2$ period (4.3$\pm$0.4 kpc and 8.7$\pm$0.9 kpc in the low and high mass bin, respectively). In both mass bins, the green histograms show  the size functions $\Phi_{MQ}(r_{1/2})$  of  star-forming discs predicted to be  mass-quenched  between $z=1$ and $z=0.2$ according to the continuity equation of Peng et al.\ (2010; P10). The green arrows show the median sizes of  these predicted mass-quenched galaxy populations  (5.8$\pm$0.1 kpc and 8.7$\pm$0.1 kpc in the low and high mass bin, respectively). \label{f8}}
\end{figure*}

\begin{figure*}
\epsscale{1.1}
\plotone{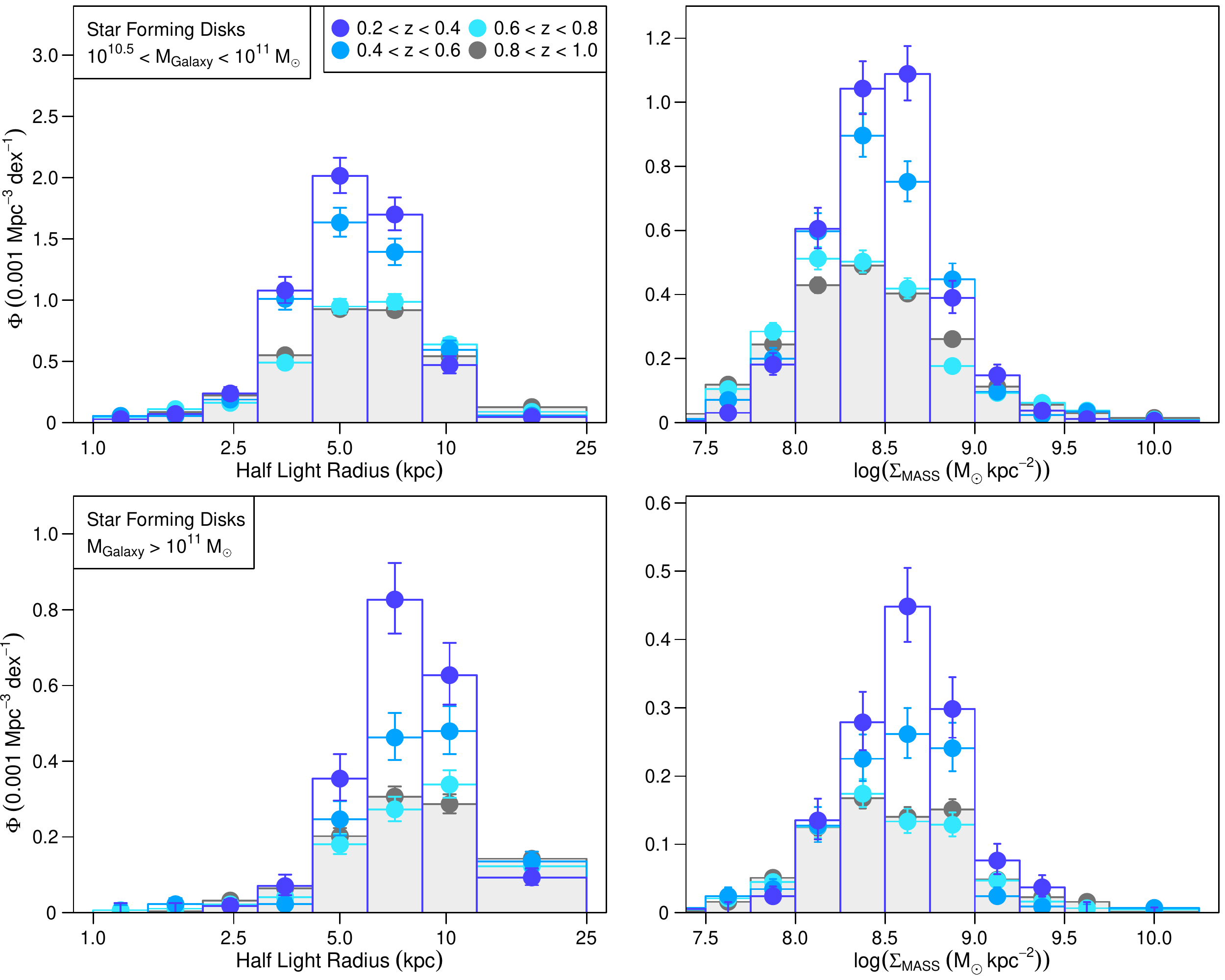}
\caption{The size (i.e., half-light radius) and surface mass density $\Sigma_\mathrm{MASS}$ distributions of star-forming disk galaxies  in COSMOS, in the redshift interval  $0.2 < z < 1$.  Results are shown for two bins of stellar mass, $10^{10.5} < M_\mathrm{Galaxy} < 10^{11}$$M_\odot$ (top row) and $M_\mathrm{Galaxy} > 10^{11}$$M_\odot$ (bottom row).  
As in Fig.\ \ref{f4}, the redshift interval is split in four bins, i.e., $0.2 < z < 0.4$ (dark blue), $0.4 < z < 0.6$ (light blue), $0.6 < z < 0.8$ (cyan), $0.8 < z < 1.0$ (grey). 
The error bars on each histogram represent only the Bayesian 1$\sigma$ CIs accounting for the Poisson noise in each bin; photometric sample incompleteness and (the non-Poissonian component of) cosmic variance contribute further uncertainties on the order of 15\% each to the absolute normalization (Section \ref{dataset}).   As in Fig.\  \ref{f4},  the normalization of the two intermediate redshift bins is corrected for the known cosmic variance in the field (see text). \label{f9}}
\end{figure*}

 We can thus finally compute the size function $\Phi_{MQ}(M_{Galaxy}, r_{1/2})$ of star-forming  galaxies of a given mass $M_{Galaxy}$  which  should be   mass-quenched and transferred to the Q-ETG sequence between $z=1$ and $z=0.2$  by summing up the  size functions of star-forming galaxies  of Fig.\ \ref{f9}  in the three highest redshift bins of our analysis, each weighted by its own mass-quenching fraction as given above, i.e.:

 \begin{equation}
\Phi_{MQ}(r_{1/2}) =  \sum_{i=1}^3  f_{MQ, i} \times \Phi_{SF, i}(r_{1/2})
\end{equation}

with $\Phi_{SF, i}$ the size function of star-forming galaxies in the $i$-th redshift bin, and $i$ running through our three highest redshift bins. The results are presented in Fig.\ \ref{f8} as green histograms; their medians are  indicated  by the green  arrows, and are equal to 5.8$\pm$0.1 kpc and 8.7$\pm$0.1 kpc in the low and high mass bins, respectively.   If mass-quenching of star-forming galaxies of similar mass is responsible for the appearance of the large-size Q-ETG populations observed in both our mass bins, and if mass-quenching does not change the sizes of the progenitors, these predicted distributions for the newly-quenched galaxies should be similar to the corresponding empirical black-hatched histograms of Fig.\ \ref{f8}.
 
 At lower masses, $10^{10.5} < M_\mathrm{Galaxy} < 10^{11} M_\odot$, the comparison between black  and green histograms readily shows   a  good agreement between the number densities integrated over all sizes  of  predicted and observed NQ-ETGs (respectively $2.1\pm0.2 \times 10^{-4} $Mpc$^{-3}$  and $2.6\pm0.2 \times 10^{-4}$ Mpc$^{-3}$).  At face value, however,  the median size predicted for    mass-quenched star-forming galaxies   is larger by about $30\%-40\%$  relative to that  derived  for  NQ-ETGs in the  $z=1\rightarrow0.2$ period. 
This observed difference in the median sizes of the newly-quenched remnants relative to their star-forming progenitors may be explained  as an `apparent' (rather than physical) shrinkage,  due to    rapid fading of the disk components of these galaxies in the aftermath of their mass-quenching episode.  This scenario is supported by the fact that, as we discuss above,  $>90\%$ of the star-forming progenitors are disk galaxies with a dominant bulge component.
 This implies that mass-quenching  efficiently retains the disk components of the star-forming progenitors, i.e., it has no direct effect of galaxy morphologies. This result is  consistent with the morphological mix of $z<1$ quenched galaxies in COSMOS (e.g., \citealt{2013MNRAS.428.1715H}; Oesch et al.\ 2010), and with the analysis of the morphological mix of the quenched  population in $z\sim0$ virialized dark matter halos, which is found in ZENS (Carollo et al.\ 2013a) to remain  constant with halo-centric distance, despite the large increase of the total fraction of quenched satellites towards the halo centers (Carollo et al.\ 2013b, in preparation). 

Also at masses  $M_\mathrm{Galaxy}>10^{11} M_\odot$   there is an excellent agreement in the number densities, integrated  between $z=1$ and $z=0.2$, of  'predicted' and 'observed' newly-quenched galaxies in each individual size bin  (and thus also in the total number densities integrated over all sizes, which are respectively  $3.2\pm0.2 \times 10^{-4} $Mpc$^{-3}$  and $3.1\pm0.5 \times 10^{-4}$ Mpc$^{-3}$).  In addition, at these high masses,   the median size  for the predicted mass-quenched population of star-forming disk galaxies  over the  $z=1\rightarrow0.2$ time span (green histogram) is strikingly  similar to that  inferred, over the same redshift range, for the newly-quenched population of Q-ETGs   (black histogram).   This means that, if also at these high masses  mass-quenching  of progenitor star-forming disk galaxies  drives   the average size evolution of the QETG population, the physical mechanism  behind mass-quenching  should  conserve stellar mass and   size, while transforming morphologies from disk-like to spheroidal-like.  At least in part, this could be explained again by fading of the  disk components of the progenitor bulge-dominated star-forming galaxies, since  at such high  masses  these  progenitors are likely to have   high bulge-to-disk ratios and thus  small disk components of sizes comparable to those of the dominant bulges. As already commented above, however, at these high masses other factors such as individual galaxy growth and  merging are also likely to contribute to the evolution of the Q-ETG population, as indicated by the  detected  $\sim30\%-40\%$ decrease towards lower redshifts in the number density of the most compact of such systems, and by  the similarity in rest-frame   colors of   small and large galaxies.

\section{Summary and conclusions}\label{lastwords}

In the previous sections, we have argued that the apparent size evolution of the Q-ETG population is largely due to the appearance of new, larger, galaxies adding to a rather stable pre-existing population of more compact galaxies which has a more or less constant number density.  We do not claim that this is the only effect in operation, but it is certainly in our view the dominant one.

This conclusion is most strongly supported in the lower mass bin of Q-ETGs that we have considered in this paper, $ 10^{10.5}  < M_{Galaxy}< 10^{11} M_\odot$, which samples the bulk of the quenched early-type population. At these masses, our data show $(i)$ no evolution in the number density  of the smallest and densest Q-ETGs between $z=1$ and $z=0.2$, and $(ii)$ evidence of a strong build up in the population of their larger siblings over the same redshift range.   In other words, at these masses, Q-ETGs roughly have a  constant size, once formed, and   newly formed Q-ETGs are being added at larger sizes.  Previous works had suggested that Q-ETGs
evolve mildly in size, once formed, and
that the evolution of the mass-size relation is also driven by newly formed Q-ETGs being added at
larger sizes (e.g., Valentinuzzi et al.\ 2010; Cassata et al.\ 2011; Newman et al.\ 2012; Poggianti et al.\ 2013). 
Our results  put on a firm foundation
and quantify these earlier suggestions, and show that, especially at sub-$M^*$ masses, it is the addition of newly formed Q-ETGs   at larger sizes  rather than the growth of individual galaxies 
that dominates  the 
evolution of the mass-size relation.

The alternative interpretation that galaxies are individually expanding in size while the new ones are added to replace them at the {\it small}-end of the size distribution runs counter to our finding that $(i)$ the compact quenched population shows increasingly redder $(U-V)$ colors towards lower redshifts, with an average color difference between $z=1$ and $z=0.2$ that is well-consistent with the aging, over this time span, of a passively-evolving galaxy population, and $(ii)$ that the rest-frame $(U-V)$ colors of the larger Q-ETG are systematically bluer (as also suggested  at the higher and lower redshifts respectively by Saracco et al.\ 2011  and van der Wel et al.\ 2009), pointing at  a more recent quenching than the corresponding compact population.  
Together this evidence excludes the possibility that the  stable number density between $z=1$ and $z=0.2$ of compact Q-ETGs could be due to a balance between the formation rate of new compact Q-ETGs and their depletion rate due to a size growth out of the compact bin, and indicates that the newly quenched galaxies are being added at the larger end of the size distribution.

The growth of the median radius of the quenched early-type population has usually been interpreted in terms of the evolution of individual galaxies (e.g., Oser et al 2012), without considering the changes in the number density that, as argued here, indicate that the effect is driven by the addition of newly-quenched larger galaxies at later epochs.  The most popular mechanism has been size-increase through minor merging (e.g., Hopkins et al.\ 2009; Feldmann et al.\ 2010; Cimatti et al.\ 2012). 
This latter scenario for the growth of the median size of Q-ETGs requires however about 10 mergers with about 1:10 mass ratios to explain the observed size growth since $z \sim 2$ (e.g. van de Sande et al. 2012  and references therein). None of these minor mergers can be gas-rich, since it is believed that gas infall towards the primary galaxy center would lead to nuclear star formation and thus to a `shrinkage' of its half-light radius.  Such a sequence of ten dry mergers with no intervening wet mergers seems unlikely, especially on account of the fact that, at the  $\sim10^{10} M_\odot$ mass scale of the 1:10 companions of $\sim10^{11} M_\odot$ galaxies, gas-rich systems largely outnumber their dry counterparts. This adds to another difficulty of this scenario, namely the dearth of suitable nearby companions highlighted by Newman et al. (2012).

 Nevertheless, it is clear that some such individual size evolution must have happened (although our data show that it is not the dominant driver of the size evolution of the population).  Not least, our own analysis has indicated a modest decrease in the number density of compact $r_{1/2}<2$ kpc galaxies at $M_{Galaxy} > 10^{11} M_\odot$ by $\sim 30\%-40\%$ over the  redshift interval $z=1 \rightarrow 0.2$.  These very massive galaxies also show weaker color trends; we highlight   here, as an open issue, to understand the similarity between the average  rest-frame  colors of compact and large quenched early-type galaxies with masses above $M^*$.

A fact to keep in mind is that quenched galaxies that are as massive as $10^{11} M_\odot$ and above already by redshift $z\sim1$  will typically be the central galaxies in haloes that have a mass today of  $\geq 10^{13} M_\odot$.   Hydrodynamical simulations  in a $\Lambda$-CDM   universe show that, over this period, such galaxies grow in stellar mass mostly through minor mergers; these mergers produce outer halos and leave the central density of galaxies almost constant (e.g., Hopkins et al 2009; Feldmann et al.\ 2010).   We also note that in the continuity analysis of Peng et al (2010), $10^{11} M_\odot$ was identified as a threshold above which post-quenching mass increase through mergers was likely to be significant, and below which it was generally unlikely to have been. The 30\% decrease in density found here, and possibly the colors, of these very massive quenched early-type galaxies 
are thus consistent with a picture in which a fraction of them accrete diffuse halos at later times, as also suggested by many other  studies (see quoted references).   We re-emphasize however that even at this high mass-end the dominant effect in the increase of the average radius for the whole population remains the appearance, at later epochs, of  equally massive but larger (in half-light size) Q-ETGs. 

The present analysis indicates that the typical sizes of newly quenched ETGs increase with time,   i.e., that quenching of progressively larger star-forming galaxies progressively outnumbers that of galaxies with smaller sizes (which dominated the quenching rate at earlier epochs); this is the dominant cause of the apparent increase in size of the integrated quiescent population.  
The typical sizes these of newly quenched galaxies appear  to scale as roughly $(1+z)^{-1}$, suggesting that the mean stellar density within these newly quenched galaxies scales as the mean density of the Universe at the time of quenching.  This idea may unify a number of important observational facts, including the observed correlation between sSFR and mean stellar density in the local Universe.

\acknowledgements{TB, AC, EC and MO thank the Swiss National Science Foundation for financial support. AR acknowledges the ETHZ Institute of Astronomy for its kind hospitality at the time when this research was started
and this paper first laid down, and the National Institute for Astrophysics for support through the grant INAF-PRIN 2010. }
\appendix

\section{A. The Choice of SED-Based SFRs}\label{sfrs}

Before adopting as  fiducial SFRs those derived from the SED fits, we analyzed in detail  SFR estimates derived using $(a)$ dust-extinction corrected restframe FUV fluxes, using the Calzetti et al.\ (2000) extinction curves together with the reddening relation $E(B-V)=0.23\beta+0.49$ of Meurer et al.\ (1999), as well as  the Wijesinghe et al. (2011)  reddening relation $E(B-V)=0.2\beta+0.33$ (with $\beta$ the slope of the UV continuum in a power-law approximation);  $(b)$ Spitzer/MIPS   $24\mu$m detections down to a limiting flux of $80\mu$Jy with  Dale \& Helou (2002) templates (MIPS catalog courtesy of E.\ Le Floch); and $(c)$ the combination of  IR-based and  FUV-based SFRs, the latter not corrected for extinction. 

Given the flux limits of the available FUV and Spitzer COSMOS photometry, the FUV- or $24\mu$m-based SFRs result in strong and redshift-dependent selection biases towards  mass and SFR. The FUV selection  picks up 75\% of galaxies at the redshifts of interest, but it misses the most massive and quiescent systems (median mass of the FUV-detected galaxies is $\log(M/M_\odot)=9.4$, compared with a median mass of the galaxies not detected in FUV of
$\log(M/M_\odot)=9.9$). Infrared-based SFRs are only available for about 17\% of the COSMOS galaxies and pick up only the most massive star-forming objects (IR median mass $<\log{M/M_\odot}=10.3$, compared to 9.3 for  galaxies not detected in the MIPS images in the $0.2< z <1.0$ range), with a rapidly increasing star formation rate detection threshold  as a function of redshift.

\begin{figure*}
\epsscale{1.1}
\plotone{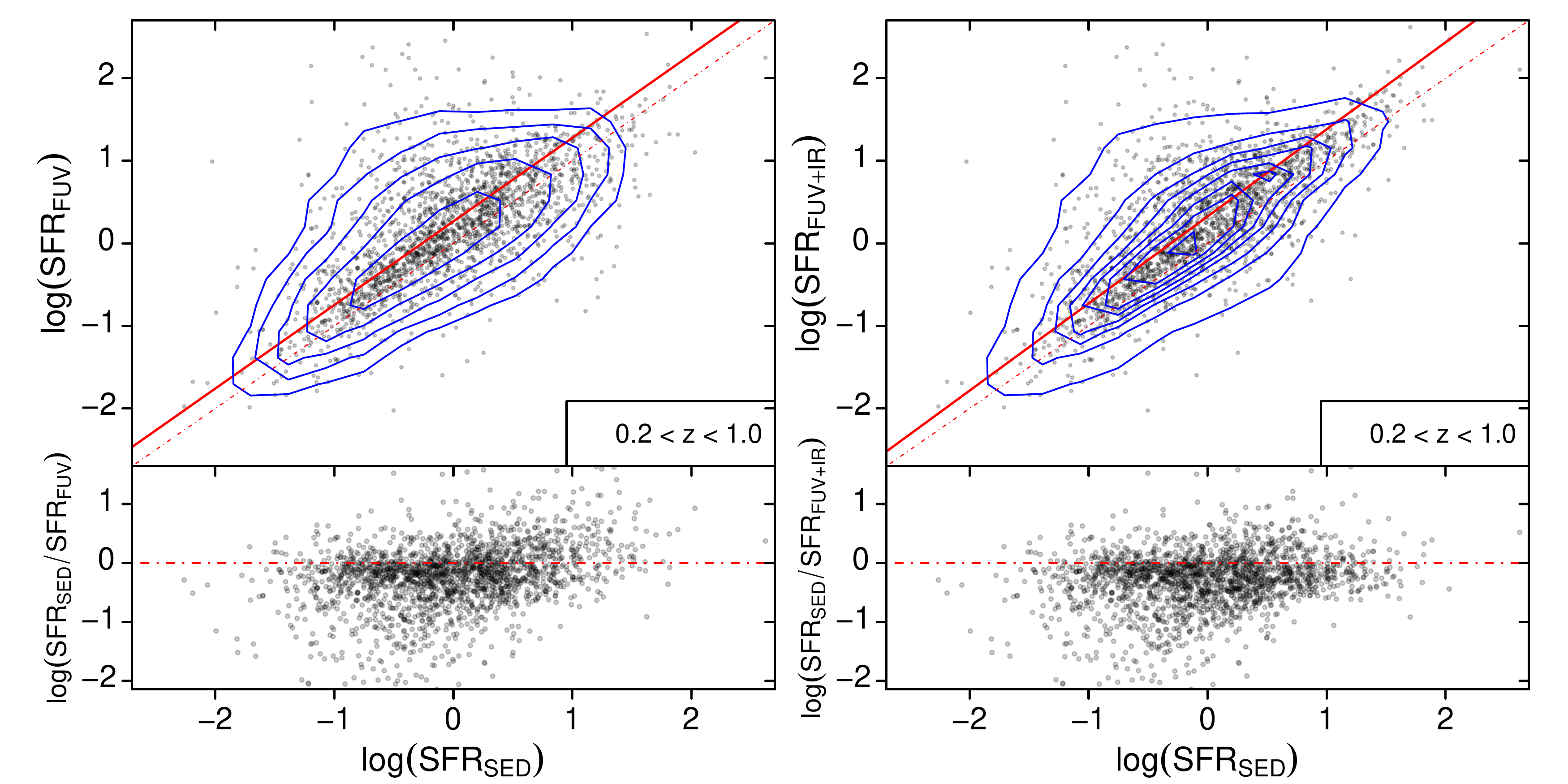}
\caption{{\it Left panel:}  Fit  to the relation between  UV-based, dust-corrected SFRs and corresponding values  derived from SED fits to galaxy photometric points. {\it Right panel:}  Fit to the relation between  the sum of IR-based and dust-uncorrected UV-based SFRs, and SFRs derived from SED fits to  galaxy photometric points. The good correlation between $SFR_{FUV,cor}$ and  $SFR_{IR+UV}$ with $SFR_{SED}$ motivates our use of the latter, which is available for the whole  $I_{814W}<24$ COSMOS sample. Note that only one out of 25 points is plotted, for clarity of presentation. The contours refer however to the whole sample (with each line referring to an increase in number density of a factor of two). \label{f10}}
\end{figure*}

The comparison of the extinction corrected FUV-based SFRs with  the SFRs derived by summing the uncorrected-FUV plus IR contributions,  favors the Meurer et al.\ reddening relation.   Comparison of the FUV SFRs, corrected for extinction using this relation, shows a small shift but, overall, an excellent agreement with the SFRs derived from our SED fits. A fit of the relation gives: $\log(SFR_{FUV,cor})=1.01\pm{0.01}\times\log(SFR_{SED})+0.27\pm{0.01}$, with a formal 1-$\sigma$ scatter of 0.48  (see left panel of Figure \ref{f10}). A similar fit to the relation between SED-based SFRs and SFRs obtained by summing IR-based and (dust-uncorrected) UV-based values gives: $\log(SFR_{FUV+IR})=1.06\pm{0.01}\times\log(SFR_{SED})+0.33\pm{0.01}$, with a formal 1-$\sigma$ scatter of 0.45  (see right  panel of Figure \ref{f10}). These excellent agreements validate our adoption  of the SED-based SFRs for our analysis, which have the virtue to be uniformly derived and available for all galaxies in our sample. We stress however  that we have tested that our results do not depend on the choice of SFR diagnostic adopted for this analysis.

\begin{figure*}[!b]
\epsscale{1.1}
\plotone{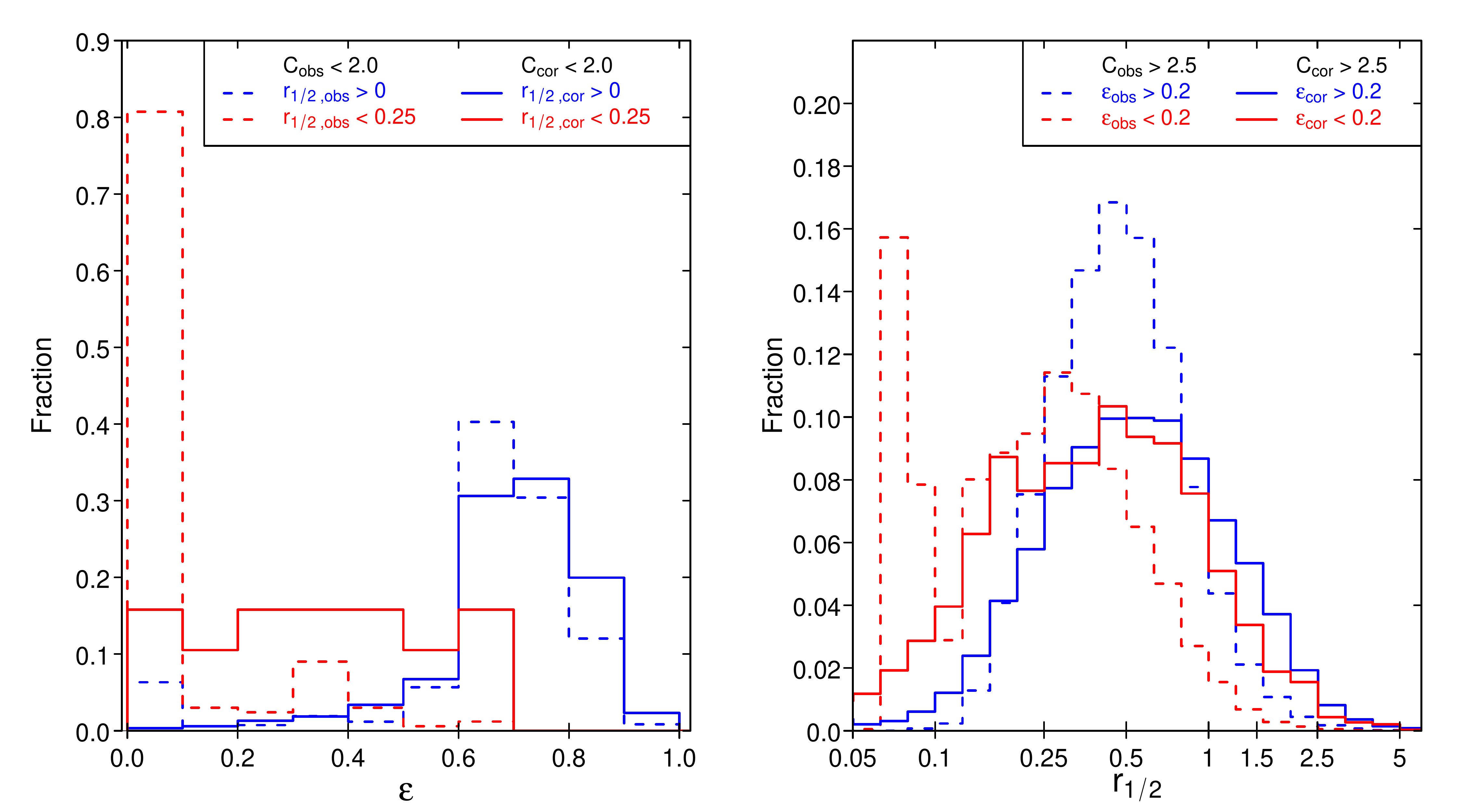}
\caption{{\it Left panel:} The uncorrected (dashed line) and corrected (solid line) distribution of ellipticities for $C<2$ disk galaxies in COSMOS. Blue is for the whole $C<2$ sample, red is for the subsample of  $r_{1/2}<0.25"$ galaxies. {\it Right panel:} The uncorrected (dashed) and corrected (solid) sizes distributions of $C>2.5$ disk galaxies in COSMOS. All histograms are normalized to the total number of galaxies in the plotted samples. Both panels show that systematics in the raw size measurements are substantially reduced after applying the corrections of Section \ref{seccor}.
 \label{f11}}
\end{figure*}

\begin{figure*}[!b]
\epsscale{1.1}
\plotone{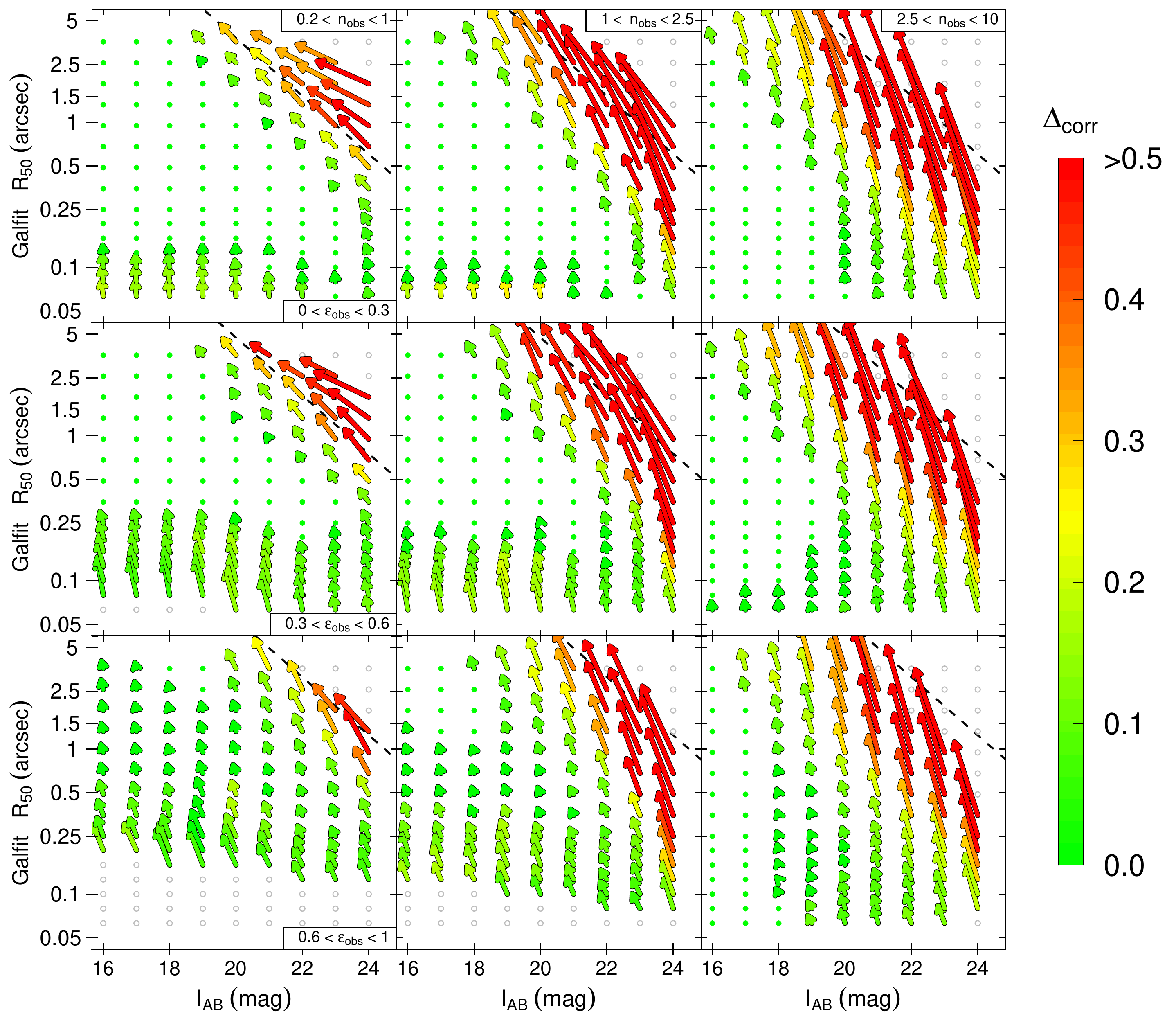}
\caption{The correction grid for {\it observed} half-light radius and magnitude values for $I_{814W}<24$ mag COSMOS galaxies, as already presented in Fig.\ ref{f2}, but this time for size and magnitude measurements obtained through Galfit fits to the galaxies' surface brightness distributions. The three bins of concentration index of Fig.\ \ref{f2} are here replaced by three bins in S\'ersic index $n$. Colors and symbols are as in Fig.\ \ref{f2}.\label{f12}}
\end{figure*}

\begin{figure*}[!t]
\epsscale{1.1}
\plotone{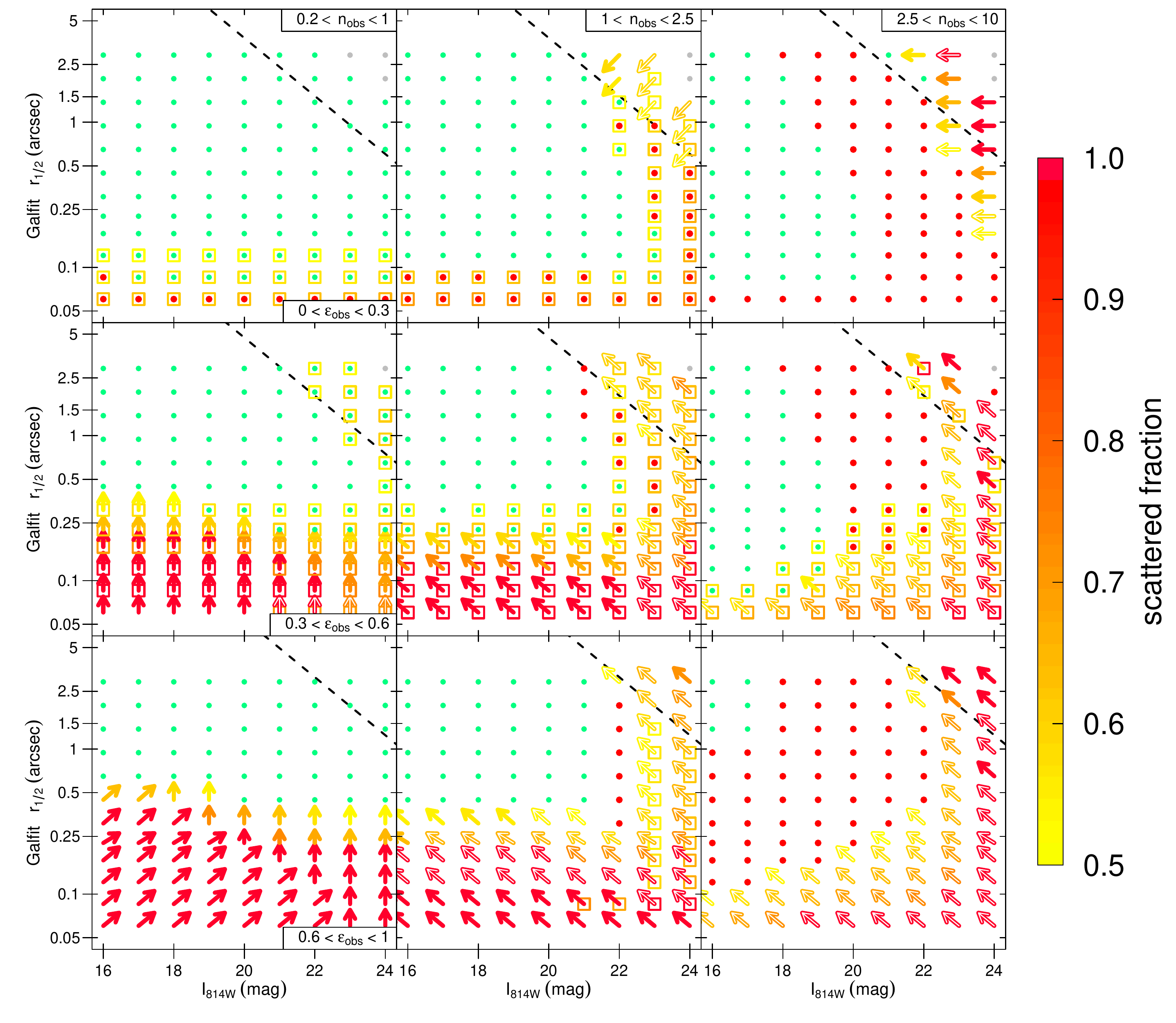}
\caption{Analysis of scattering of model galaxies in- and out- of true/input ellipticity and concentration bins, as in Fig.\ ref{f3}, but this time for measurements obtained through Galfit fits to the galaxies' surface brightness distributions. Similarly to  Fig.\ \ref{f12}, the three bins of concentration index of Fig.\ \ref{f3} are here replaced by three bins in S\'ersic index $n$. Colors and symbols are as in Fig.\ \ref{f3}.\label{f13}}
\end{figure*}

\section{B. Reliability of the Correction Functions for Galaxy Sizes:  \\ Consistency Checks}\label{testscorrs} 

Simple analyses and considerations using the uncorrected  and corrected   (with the corrections of Section \ref{seccor})  magnitude-size distributions shows the importance of accounting for systematic biases in Kron-style size measurements.  

 The left panel of figure  \ref{f11} shows the distributions of uncorrected and corrected ellipticities of galaxies with concentration values $C<2$, for the entire sample (blue) and the sample limited by $r_{1/2}<0.25"$.  Uncorrected  measurements  would lead to an unphysical dependence of sizes of {\it small}  disk on inclination/viewing angle, as evident from the figure.  The correction functions restore, from the low ellipticity bin to the high ellipticity bin, a fraction of intrinsically small,  edge-on disks that were rounded/blurred by the observational PSF, and furthermore recover their intrinsic sizes, which were reduced by  this PSF blurring.  
 The right panel in the same figure shows the uncorrected and corrected size distributions of $C> 2.5$ galaxies, mostly spheroids and bulge-dominated disks, in two bins of ellipticity, above and below $\epsilon=0.2$.  Results are again shown in red and blue for uncorrected and corrected quantities, respectively. A significant fraction of intrinsically high ellipticity, $C>2.5$ galaxies which our morphological analysis and inspection confirm to be primarily edge-on, bulge-dominated disks, are  returned by the correction functions  to the $e_\mathrm{cor} > 0.2$ bin. Furthermore, the size distributions of round and inclined systems are in much better agreement after the application of the corrections.

\section{C. The Need for Correcting Sizes Derived from Surface Brightness Fits: \\ A Showcase with Galfit}\label{galfitsizes}

As highlighted in Section \ref{sizecorrections} it is not only (Kron, Petrosian, or isophotal) aperture-based measurements which require a correction function to aid the recovery of true galaxy magnitudes, sizes, ellipticities, and concentration indices from their raw (i.e., observationally biased) counterparts.  In this Appendix we illustrate the nature of these biases for the case of S\'ersic (1963)  profile\footnote{The well-known S\'sersic radial profile is defined as 
$\mu(r) = \mu_{0} exp[-(r/\alpha)^{1/n}]$, where $\mu(r)$ is the intensity at radius $r$ and $\mu_{0}$ is the intensity at the galaxy center, i.e., in $r=0$. $\alpha$ is the scalelength, i.e., the radius at which the intensity drops by a factor $e^{-1}$.  The parameter $n$ is the 'S\'ersic index', and controls the degree of central concentration, i.e., the  shape  of the profile.}  modeling with Galfit (Peng et al.\ 2002) using a process of artificial galaxy simulation equivalent to that introduced for the calibration of the SExtractor/{\it ZEST+} measurements that we adopt in out analysis.   Specifically, we begin by generating $>$800,000 artificial galaxy images with input parameters randomly drawn within the following limits:  $19 < I_{814W} < 25$ mag, $0.05 < r_{1/2} < 2.5$ arcsec, $0 < e < 0.95$, and S\'ersic  index $0.2 < n < 8$.  As before each model galaxy was constructed at a pixel scale of 0.03 arcsec and convolved with a representative ACS $I_{814W}$ PSF \citep{rho07}, and added with Poisson noise to an empty region of sky from the real COSMOS HST/ACS imaging.  SExtractor was then run to generate object masks and to estimate starting values for use in Galfit.  In particular, we adopt the Kron magnitude, the mean isophotal ellipticity and position angle, and the Kron half light radius as initial guesses for their model counterparts---and all fits were begun with an initial guess of $n=2.5$ for the S\'ersic index.  After running Galfit the output parameters of all convergent fits (i.e., those which did not end in a catastrophic error causing the program to crash) were saved for comparison against our input model parameters.  The results of this analysis are presented in Figs.\ \ref{f12} and \ref{f13} (analogs to Figs.\ \ref{f2} and \ref{f3}, respectively).

Examination of the 'correction vectors' diagram for the Galfit simulations shows two  important differences relative to the analog diagram for the SExtractor/{\it ZEST+} case shown in Fig.\ \ref{f2}: \textsc{(i)} not surprisingly, the incorporation of the observational PSF into the fitting algorithm Galfit does indeed outperform SExtractor/{\it ZEST+} in size recovery for the most compact galaxies; however, \textsc{(ii)} in the low surface brightness regime,  the Galfit-based measurement are substantially affected by systematic offsets from their true (input) values---with a clear trend  towards under-estimation of both apparent magnitudes and sizes.  Note also that, as indicated by the color coding of the error vectors in Figs.\  \ref{f12} and \ref{f13}, the random errors in size recovery of  high concentration/n-S\'ersic systems with Galfit are worse in the low surface brightness regime than they are for the more stable Kron approach (as implemented in SExtractor/ZEST+). Corrections to  model-based measurements  may be important especially at high redshifts, where faint disk components may populate  the low surface brightness regime of the correction diagrams.  

\begin{figure}
\epsscale{1.1}
\plotone{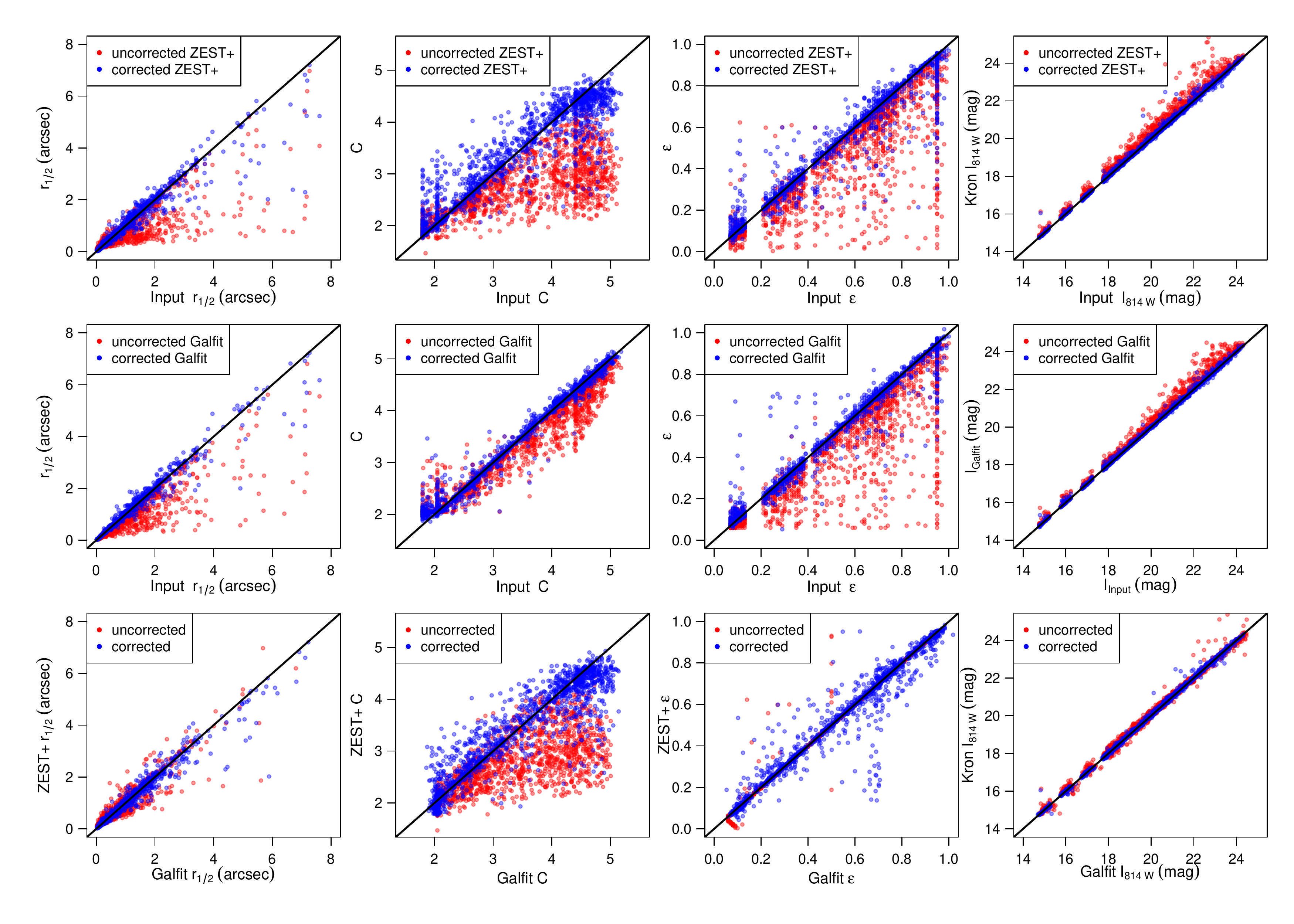}
\caption{Comparison between uncorrected (red points) and corrected (blue points) parameters obtained with ZEST+ (top row) and Galfit (middle row). Corrected parameters are obtained by applying the calibrations discussed in Section \ref{sizecorrections}. The shown parameters are, from left to right: half-light radius $r_{1/2}$, concentration C, ellipticity $\epsilon$, and I$_{814W}$ magnitude. For clarity, the plot shows results for only 1500 of the simulated galaxies.  
The bottom row shows the comparison between measurements obtained with ZEST+ and Galfit before (red) and after (blue) the application of the corrections.  The applied calibrations substantially improve the scatter between "true" (input) and "observed" values, and lead to a very good agreement between {\it corrected} measurements obtained with either the ZEST+ aperture approach or the Galfit surface-brightness fitting approach.  \label{f14}}
\end{figure}

Once the Galfit and Kron-based  size measurements  are properly corrected for, as we do in this paper, then both estimates obtained with either SExtractor/{\it ZEST+} or Galfit are in excellent agreement, as shown in   Fig.\ \ref{f14}.  The figure shows the inherent consistency of our correction functions through a comparison of the apparent magnitudes and sizes of thirty  randomly drawn high Sersic-$n$ COSMOS galaxies,  recovered after applying our correction maps to both the SExtractor/{\it ZEST+} and the Galfit measurements.  Although both measurement techniques require different and, in both cases, non-negligible adjustments against systematic recovery biases in this parameter space, the resulting distributions are in very good agreement, with only small random offsets between the two sets of estimates.  

This gives us confidence that the  results of this study are robust against the choice of size measurement algorithm owing to the use of our correction function approach.  

\section{D. The quenched fractions from Peng et al.\ 2010}\label{peng}

To compare the Peng et al.\ predictions with our estimates, we convert the mass and time dependent evolution of the  sSFR in their Eq.\ (1) :

\begin{equation}
 \mathrm{sSFR(M_{Galaxy},t)}=2.5\left( \frac{M_{Galaxy}}{10^{10}M_\odot}\right)^\beta \left( \frac{t}{3.5 \mathrm{Gyr}}\right)^{-2.2}\mathrm{Gyr^{-1}}
 \end{equation}
 
(with $\beta = -0.1$, see references above),  into the mass and time dependent evolution of the  {\it reduced} specific star--formation rate, $rsSFR(M_{Galaxy},t)$, as done in Lilly et al.\ (2013, submitted to ApJ); this is necessary   since, in contrast with those authors (who use the 'actual' stellar mass trapped in long-lived stars in their study), we defined the stellar mass as equal to the integral of the SFR (see Section \ref{stellarmasses}). We have:

\begin{equation} \label{rsSFR}
rsSFR(M_{Galaxy},t) = (1-R) \times sSFR,
\end{equation}

where $R$ is the fraction of the baryonic mass that is converted into stars and is promptly returned to the interstellar medium (with the remaining fraction $1-R$ remaining locked  in long-lived stars).  The inverse of  the $rsSFR$ sets the  characteristic timescale for the build-up of the long-lived stellar population.  As in  Lilly et al.\ 2013, we assume  an instantaneous mass return and  $R = 0.4$, as derived from stellar population models of Bruzual \& Charlot (2003). 

The fraction $f_{MQ}$ of  transient, just Òbeing-quenchedÓ objects  at any time $t$
 can be derived from Eq. (27) of \citet{pen10}, modified to take into account Eq.\ \ref{rsSFR} above, by computing:

 \begin{equation} 
f_{MQ}(t) = \frac{\mathrm{\Phi_{SF}(t)} \times rsSFR(M_{Galaxy},t) \times \frac{M_{Galaxy}}{M^*}}
{\mathrm{\Phi_{SF}(t)}}, 
\end{equation}
 
with $\Phi_{SF}(t)$ the number density of star-forming galaxies at the same epoch. We can thus predict the fractions $f_{MQ, i}$ of newly-quenched galaxies in each redshift bin $i$ of our analysis as:

\begin{equation}
f_{MQ, i} = f_{MQ}|_{t(z_{high, i})}^{t(z_{low, i})}=\frac{M_{Galaxy}}{M^*}\int_{t(z_{high, i})}^{t(z_{low, i})}{\mathrm{rsSFR(M_{Galaxy},t)} dt}, 
\end{equation}

\vspace*{0.2truecm} 
with $z_{high, i}$ and $z_{low, i}$ the upper and lower limit of  the $i$-th redshift bin.

\begin{thebibliography}{}

\bibitem[Abazajian et al.(2009)]{abazajian_et_al_2009} Abazajian, K.~N., 
Adelman-McCarthy, J.~K., Ag{\"u}eros, M.~A., et al.\ 2009, \apjs, 182, 543 
\bibitem[Barro et al.(2013)]{2013ApJ...765..104B} Barro, G., Faber, S.~M., 
P{\'e}rez-Gonz{\'a}lez, P.~G., et al.\ 2013, \apj, 765, 104 
\bibitem[Bertin 
\& Arnouts(1996)]{1996A&AS..117..393B} Bertin, E., \& Arnouts, S.\ 1996, \aaps, 117, 393 
\bibitem[\protect\citeauthoryear{Brammer et al.}{2009}]{bra09}Brammer, G. B., 2009, ApJ, 706, 1, L173
\bibitem[\protect\citeauthoryear{Bruzual \& Charlot}{2003}]{bru03}Bruzual, G., Charlot, S., 2003, MNRAS, 344, 4, 1000
\bibitem[\protect\citeauthoryear{Bundy et al.}{2010}]{bun10}Bundy, K., et al., 2010, ApJ, 719, 2, 1969
\bibitem[Calzetti et al.(2000)]{2000ApJ...533..682C} Calzetti, D., Armus, 
L., Bohlin, R.~C., et al.\ 2000, \apj, 533, 682 
\bibitem[\protect\citeauthoryear{Cameron \& Driver}{2007}]{cam07}Cameron, E., Driver, S. P., 2007, MNRAS, 377, 1, 523
\bibitem[\protect\citeauthoryear{Cameron et al.}{2010}]{cam10}Cameron, E., et al., 2010, MNRAS, 409, 1, 346
\bibitem[\protect\citeauthoryear{Capak et al.}{2007}]{cap07}Capak, P., et al., 2007, ApJS, 172, 99
\bibitem[\protect\citeauthoryear{Cardelli et al.}{1989}]{car89}Cardelli, J. A., Clayton, G. C., Mathis, J. S., 1989, ApJ, 1, 345, 245
\bibitem[Carollo et al.(2012)]{2012arXiv1206.5807C} Carollo, C.~M., Cibinel, A., Lilly, S.~J., et al.\ 2013a, arXiv:1206.5807 
\bibitem[Cassata et al.(2011)]{2011ApJ...743...96C} Cassata, P., 
Giavalisco, M., Guo, Y., et al.\ 2011, \apj, 743, 96 
\bibitem[\protect\citeauthoryear{Chabrier}{2003}]{cha03}Chabrier, G., 2003, ApJ, 586, 2, L133
\bibitem[Cibinel et al.(2012)]{2012arXiv1206.6496C} Cibinel, A., Carollo, C.~M., Lilly, S.~J., et al.\ 2013, arXiv:1206.6496
\bibitem[Cimatti et al.(2006)]{2006A&A...453L..29C} Cimatti, A., Daddi, E., \& Renzini, A.\ 2006, \aap, 453, L29
\bibitem[Cimatti et al.(2008)]{2008A&A...482...21C} Cimatti, A., Cassata, P., Pozzetti, L., et al.\ 2008, \aap, 482, 21
\bibitem[Cimatti et al.(2012)]{2012MNRAS.422L..62C} Cimatti, A., Nipoti, 
C., \& Cassata, P.\ 2012, \mnras, 422, L62 
\bibitem[\protect\citeauthoryear{Coleman et al.}{1980}]{col80}Coleman, G. D., Wu, C.-C., Weedman, D. W., 1980, ApJS, 43, 393
\bibitem[Covington et al.(2011)]{2011MNRAS.415.3135C} Covington, M.~D., Primack, J.~R., Porter, L.~A., et al.\ 2011, \mnras, 415, 3135
\bibitem[Daddi et al.(2005)]{2005ApJ...626..680D} Daddi, E., Renzini, A., Pirzkal, N., et al.\ 2005, \apj, 626, 680 
\bibitem[Dale 
\& Helou(2002)]{2002ApJ...576..159D} Dale, D.~A., \& Helou, G.\ 2002, \apj, 576, 159 
\bibitem[Dom{\'{\i}}nguez S{\'a}nchez et al.(2011)]{2011MNRAS.417..900D} Dom{\'{\i}}nguez S{\'a}nchez, H., Pozzi, F., Gruppioni, C., et al.\ 2011, \mnras, 417, 900
\bibitem[Dullo 
\& Graham(2013)]{2013ApJ...768...36D} Dullo, B.~T., \& Graham, A.~W.\ 2013, \apj, 768, 36 
\bibitem[\protect\citeauthoryear{Elmegreen et al.}{2009}]{elm09}Elmegreen, B. G., Elmegreen, D. M., Fernandez, M. X., Lemonias, J. J., 2009, ApJ, 692, 1, 12
\bibitem[\protect\citeauthoryear{Feldmann et al.}{2006}]{fel06}Feldmann, R., et al., 2006, MNRAS, 372, 2, 565
\bibitem[Feldmann et al.(2010)]{2010ApJ...709..218F} Feldmann, R., Carollo, C.~M., Mayer, L., et al.\ 2010, \apj, 709, 218 
\bibitem[Franx et al.(2008)]{2008ApJ...688..770F} Franx, M., van Dokkum, 
P.~G., Schreiber, N.~M.~F., et al.\ 2008, \apj, 688, 770 
\bibitem[Giavalisco et al.(2004)]{2004ApJ...600L..93G} Giavalisco, M., Ferguson, H.~C., Koekemoer, A.~M., et al.\ 2004, \apjl, 600, L93
\bibitem[\protect\citeauthoryear{Graham et al.}{2005}]{gra05}Graham, A. W., Driver, S. P., Petrosian, V., Conselice, C. J., Bershady, M. A., Crawford, S. M., Goto, T., 2005, AJ, 130,  4, 1535
\bibitem[Guo et al.(2011)]{2011ApJ...735...18G} Guo, Y., Giavalisco, M., 
Cassata, P., et al.\ 2011, \apj, 735, 18 
\bibitem[\protect\citeauthoryear{H\"aussler et al.}{2007}]{hau07}H\"aussler, B., et al., 2007, ApJS, 172, 615
\bibitem[\protect\citeauthoryear{Hogg et al.}{2010}]{hog10}Hogg, D. W., Bovy, J., Lang, D., 2010, preprint(arXiv:1008.4686)
\bibitem[Hopkins et al.(2009)]{2009MNRAS.398..898H} Hopkins, P.~F., Bundy, K., Murray, N., et al.\ 2009, \mnras, 398, 898 
\bibitem[Hopkins et al.(2010)]{2010MNRAS.401.1099H} Hopkins, P.~F., Bundy, K., Hernquist, L., Wuyts, S., \& Cox, T.~J.\ 2010, \mnras, 401, 1099 
\bibitem[Huertas-Company et al.(2013)]{2013MNRAS.428.1715H} 
Huertas-Company, M., Mei, S., Shankar, F., et al.\ 2013, \mnras, 428, 1715 
\bibitem[\protect\citeauthoryear{Ilbert et al.}{2009}]{ilb09}Ilbert, O., et al., 2009, ApJ, 690, 2, 1236
\bibitem[Ilbert et al.(2010)]{ilb10} Ilbert, O., Salvato, M., Le Floc'h, E., et al.\ 2010, \apj, 709, 644  
\bibitem[Ilbert et al.(2013)]{2013arXiv1301.3157I} Ilbert, O., McCracken, H.~J., Le Fevre, O., et al.\ 2013, arXiv:1301.3157
\bibitem[\protect\citeauthoryear{Kinney et al.}{1996}]{kin96}Kinney, A. L., Calzetti, D., Bohlin, R. C., McQuade, K., Storchi-Bergmann, T., Schmitt, H. R., 1996, ApJ, 467, 38
\bibitem[\protect\citeauthoryear{Koekemoer et al.}{2007}]{koe07}Koekemoer, A. M., et al., 2007, ApJS, 172, 196
\bibitem[\protect\citeauthoryear{Kraft et al.}{1991}]{kra91}Kraft, R. P., Burrows, D. N., Nousek, J. A., 1991, ApJ, 374, 344
\bibitem[Kron(1980)]{1980ApJS...43..305K} Kron, R.~G.\ 1980, \apjs, 43, 305 
\bibitem[\protect\citeauthoryear{Leauthaud et al.}{2007}]{lea07}Leauthaud, A., et al., 2007, ApJS, 172, 219
\bibitem[\protect\citeauthoryear{Lilly et al.}{2007}]{lil07}Lilly, S. J., et al., 2007, ApJS, 172, 1, 70
\bibitem[\protect\citeauthoryear{Lilly et al.}{2009}]{lil09}Lilly, S. J., et al., 2009, ApJS, 184, 2, 218
\bibitem[Mancini et al.(2010)]{2010MNRAS.401..933M} Mancini, C., Daddi, E., Renzini, A., et al.\ 2010, \mnras, 401, 933
\bibitem[\protect\citeauthoryear{McCracken et al.}{2010}]{mcc10}McCracken, H. J., et al., 2010, ApJ, 708, 1, 202
\bibitem[McGrath et al.(2008)]{2008ApJ...682..303M} McGrath, E.~J., Stockton, A., Canalizo, G., Iye, M., \& Maihara, T.\ 2008, \apj, 682, 303 
\bibitem[Meurer et al.(1999)]{1999ApJ...521...64M} Meurer, G.~R., Heckman, 
T.~M., \& Calzetti, D.\ 1999, \apj, 521, 64 
\bibitem[Naab et al.(2009)]{2009ApJ...699L.178N} Naab, T., Johansson, 
P.~H., \& Ostriker, J.~P.\ 2009, \apjl, 699, L178 
\bibitem[Newman et al.(2012)]{2012ApJ...746..162N} Newman, A.~B., Ellis, R.~S., Bundy, K., \& Treu, T.\ 2012, \apj, 746, 162 
\bibitem[Nipoti et al.(2009)]{2009ApJ...706L..86N} Nipoti, C., Treu, T., Auger, M.~W., \& Bolton, A.~S.\ 2009, \apjl, 706, L86 
\bibitem[\protect\citeauthoryear{Oesch et al.}{2010}]{oes10}Oesch, P. A., et al., 2010, ApJ, 714, 1, L47
\bibitem[Onodera et al.(2010)]{2010ApJ...715L...6O} Onodera, M., Daddi, E., 
Gobat, R., et al.\ 2010, \apjl, 715, L6 
\bibitem[Oser et al.(2012)]{2012ApJ...744...63O} Oser, L., Naab, T., Ostriker, J.~P., \& Johansson, P.~H.\ 2012, \apj, 744, 63
\bibitem[\protect\citeauthoryear{Peng et al.}{2002}]{pen02}Peng, C. Y., Ho, L. C., Impey, C. D., Rix, H.-W., 2002, AJ, 124, 1, 266
\bibitem[\protect\citeauthoryear{Peng et al.}{2010}]{pen10}Peng, Y., et al., 2010, ApJ, 721, 1, 193
\bibitem[Peng et al.(2012)]{2012ApJ...757....4P} Peng, Y.-j., Lilly, S.~J., Renzini, A., \& Carollo, M.\ 2012, \apj, 757, 4
\bibitem[Poggianti et al.(2013)]{2013ApJ...762...77P} Poggianti, B.~M., Calvi, R., Bindoni, D., et al.\ 2013, \apj, 762, 77 
\bibitem[Pozzetti et al.(2010)]{2010A&A...523A..13P} Pozzetti, L., Bolzonella, M., Zucca, E., et al.\ 2010, \aap, 523, A13
\bibitem[\protect\citeauthoryear{Rhodes et al.}{2007}]{rho07}Rhodes, J. D., et al., 2007, ApJS, 172, 1, 203
\bibitem[\protect\citeauthoryear{Sanders et al.}{2007}]{san07}Sanders, D. B., et al., 2007, ApJS, 172, 86
\bibitem[Saracco et al.(2010)]{2010MNRAS.408L..21S} Saracco, P., Longhetti, M., \& Gargiulo, A.\ 2010, \mnras, 408, L21 
\bibitem[Saracco et al.(2011)]{saracco} Saracco, P., Longhetti, M., \& Gargiulo, A.\ 2011, \mnras, 412, 2707
\bibitem[\protect\citeauthoryear{Sargent et al.}{2007}]{sar07}Sargent, M. T., et al., 2007, ApJS, 172, 1, 434
\bibitem[\protect\citeauthoryear{Scarlata et al.}{2007a}]{sca07a}Scarlata, C., et al., 2007a, ApJS, 172, 1, 406
\bibitem[\protect\citeauthoryear{Scarlata et al.}{2007b}]{sca07b}Scarlata, C., et al., 2007b, ApJS, 172, 1, 494
\bibitem[Shankar 
\& Bernardi(2009)]{2009MNRAS.396L..76S} Shankar, F., \& Bernardi, M.\ 2009, \mnras, 396, L76 
\bibitem[Shankar et al.(2013)]{2013MNRAS.428..109S} Shankar, F., Marulli, 
F., Bernardi, M., et al.\ 2013, \mnras, 428, 109 
\bibitem[Schechter(1976)]{1976ApJ...203..297S} Schechter, P.\ 1976, \apj, 203, 297
\bibitem[\protect\citeauthoryear{Schlegel et al.}{1998}]{sch98}Schlegel, D. J., Finkbeiner, D. P., Davis, M., 1998, ApJ, 500, 525
\bibitem[\protect\citeauthoryear{Scoville et al.}{2007}]{sco07}Scoville, N., et al., 2007, ApJS, 172, 38
\bibitem[Scranton et al.(2002)]{2002ApJ...579...48S} Scranton, R., Johnston, D., Dodelson, S., et al.\ 2002, \apj, 579, 48 
\bibitem[S{\'e}rsic(1963)]{1963BAAA....6...41S} S{\'e}rsic, J.~L.\ 1963, 
Boletin de la Asociacion Argentina de Astronomia La Plata Argentina, 6, 41 
\bibitem[Szomoru et al.(2011)]{2011ApJ...735L..22S} Szomoru, D., Franx, M., Bouwens, R.~J., et al.\ 2011, \apjl, 735, L22 
\bibitem[\protect\citeauthoryear{Taylor et al.}{2010}]{tay10}Taylor, E. N., Franx, M., Brinchmann, J., van der Wel, A., van Dokkum, P. G., 2010, ApJ, 722, 1, 1
\bibitem[Toft et al.(2007)]{2007ApJ...671..285T} Toft, S., van Dokkum, P., 
Franx, M., et al.\ 2007, \apj, 671, 285 
\bibitem[\protect\citeauthoryear{Trenti \& Stiavelli}{2008}]{tre08}Trenti, M., Stiavelli, M., 2008, ApJ, 676, 767
\bibitem[\protect\citeauthoryear{Trujillo et al.}{2007}]{tru07}Trujillo, I., Conselice, C. J., Bundy, K., Cooper, M. C., Eisenhardt, P., Ellis, R. S., 2007, MNRAS, 382, 1, 109
\bibitem[Valentinuzzi et al.(2010)]{2010ApJ...721L..19V} Valentinuzzi, T., Poggianti, B.~M., Saglia, R.~P., et al.\ 2010, \apjl, 721, L19
\bibitem[van de Sande et al.(2012)]{2012arXiv1211.3424V} van de Sande, J., 
Kriek, M., Franx, M., et al.\ 2012, arXiv:1211.3424 
\bibitem[van Dokkum et al.(2008)]{2008ApJ...677L...5V} van Dokkum, P.~G., Franx, M., Kriek, M., et al.\ 2008, \apjl, 677, L5
\bibitem[van der Wel et al.(2009)]{vanderWel} van der Wel, A., Bell, E.~F., van den Bosch, F.~C., Gallazzi, A., \& Rix, H.-W.\ 2009, \apj, 698, 1232
\bibitem[Whitaker et al.(2011)]{2011ApJ...735...86W} Whitaker, K.~E., 
Labb{\'e}, I., van Dokkum, P.~G., et al.\ 2011, \apj, 735, 86 
\bibitem[Wijesinghe et al.(2011)]{2011MNRAS.415.1002W} Wijesinghe, D.~B., 
da Cunha, E., Hopkins, A.~M., et al.\ 2011, \mnras, 415, 1002 
\bibitem[Williams et al.(2009)]{2009ApJ...691.1879W} Williams, R.~J., Quadri, R.~F., Franx, M., van Dokkum, P., \& Labb{\'e}, I.\ 2009, \apj, 691, 1879
\bibitem[\protect\citeauthoryear{York et al.}{2007}]{yor07}York, D., G., et al., 2007, AJ, 120, 1579
\end{thebibliography}
\end{document}